\documentclass{mn2e}

\newcommand{\ltsimeq}{\raisebox{-0.6ex}{$\,\stackrel
{\raisebox{-.2ex}{$\textstyle <$}}{\sim}\,$}}
\newcommand{\gtsimeq}{\raisebox{-0.6ex}{$\,\stackrel
{\raisebox{-.2ex}{$\textstyle >$}}{\sim}\,$}}

\begin{document}

\title{A deep survey of brown dwarfs in Orion with Gemini}
\author[Lucas, Roche \& Tamura]
{P.W.Lucas$^{1}$, P.F.Roche$^{2}$ and M.Tamura$^3$\\
$^1$Dept. of Physical Sciences, University of Hertfordshire, College Lane,
Hatfield AL10 9AB.\\ email: pwl@star.herts.ac.uk\\
$^2$Astrophysics, Physics Dept., University of Oxford, 1 Keble Road, Oxford OX1 3RH.\\
$^3$National Astronomical Observatory of Japan, 2-21-1, Osawa, Mitaka, Tokyo
181-8588, Japan} 

\maketitle
\label{firstpage}

\begin{abstract}
We report the results of a deep near infrared (JHK) survey of the outer
parts of the Trapezium Cluster with Gemini South/Flamingos. 
396 sources were detected in a 26~arcmin$^2$ area, including 138 brown dwarf 
candidates, defined as M$<0.075$M$_{\odot}$ for an assumed age of 1~Myr. 
Only 33 of the brown dwarf candidates are planetary mass candidates (PMCs) 
with estimated masses in the range $0.003<$M$<0.012$M$_{\odot}$. In an extinction
limited sample (A$_V<5$) complete to approximately 0.005M$_{\odot}$ (5~M$_{Jup}$)
the mass function appears to drop by a factor of 2 at the deuterium burning 
threshold, i.e. at planetary masses. After allowing for background contamination 
it is likely that planetary mass objects at 3-13~M$_{Jup}$ number $<10\%$ of the 
cluster population, with an upper limit of 13\%. 
Analysis of the spatial distribution of stars and brown dwarf candidates 
suggests that brown dwarfs and very low mass stars (M$<0.1$M$_{\odot}$) are less 
likely than more massive stars to have wide ($>150$~AU) binary companions. This 
result has modest statistical significance (96\%) in our data but is supported at 
93\% confidence by analysis of an completely
independent sample taken from the Subaru data of Kaifu et al.(2000). There is a 
statistically very significant excess of both stars and brown dwarfs with small
separations from each other ($<6$ arcsec or 2600~AU). This appears to be due
to the presence of small N subgroups, which are likely to be dynamically unstable 
in the long term. Hence these results are consistent with the 'ejected stellar 
embryo' hypothesis for brown dwarf formation (Reipurth \& Clarke 2001). We also
report the discovery of two new bipolar nebulae, which are interpreted as Class~I 
protostars.
\end{abstract}

\begin{keywords}
stars: low mass, brown dwarfs; stars: formation; (stars:) circumstellar matter
\end{keywords}

\section{Introduction}

In recent years several deep near infrared surveys have begun to characterise
the population of brown dwarfs in the Trapezium Cluster (Lucas \& Roche 2000,
hereafter LR00; Hillenbrand \& Carpenter 2000; Luhman et al. 2000; Kaifu et al.
2000; Muench et al. 2002; Lada et al. 2004). The first three of these surveys 
were sensitive enough to detect a small population of objects whose masses 
apparently lie 
below the deuterium burning threshold at M$\ltsimeq 0.012$M$_{\odot}$
(13~M$_{Jup}$), which we call planetary mass candidates (PMCs). (Masses were 
estimated from source luminosity by comparison with pre-main sequence 
isochrones, under the assumption of a canonical age of 1~Myr). LR00 argued 
that the luminosity function declines below the deuterium threshold but the 
survey was not sensitive enough to establish this conclusively.

At least 3 deeper infared surveys with 8-m class telescopes have now been undertaken
to investigate the luminosity function and the initial mass function
to below the theoretical lower mass limit for star formation at around 
M$=6$M$_{Jup}$. This limit is based on the calculation that contracting
cloud cores with lower mass do not exceed the thermal Jeans mass at the point 
when they become optically thick to their own far infrared cooling 
radiation (Silk 1977; Low \& Lynden Bell 1973; Rees 1976). In the spherical
case it is supported by recent more detailed calculations (Masunaga \& Inutsuka 1999)
but Boyd \& Whitworth (2005) consider fragmentation of a shock compressed layer
and suggest that the mass limit may be below 3~M$_{Jup}$.
The 3 new independent surveys complement each other by using different 
combinations of survey depth and area. They will be further complemented by a
very deep optical survey currently in progress with the Hubble Space Telescope
(Robberto et al. 2004). The largest ground based survey used the VLT (oral 
presentation by M.McCaughrean at IAU Symposium 211 'Brown Dwarfs') and covers a 
region of some 50~arcmin$^2$ centred on the core of the cluster, with integration 
times of 15 minutes in each of the J, H and K near infrared filters. Our survey 
with Gemini South, which we report here,
covers an area of $\sim 26$~arcmin$^2$ in the outer parts of the cluster,
where the nebulosity is fainter and less structured. The longer integration 
times of 1 to 2 hours (see section 2) and the fainter nebulosity
increase the survey sensitivity; the fainter nebulosity also makes spectroscopic 
follow up of PMCs easier. The drawbacks are the introduction of some possible 
biases into the survey, eg. due to dynamical mass segregation between the core and the 
outer cluster, and a smaller sample for statistical analysis. The third survey,
which is still in progress (Aspin \& Liu private com.), used Gemini North to 
image a much smaller area
of $\sim 4$~arcmin$^2$ at greater depth, using near infrared filters sensitive
to methane absorption. This survey aims to pick up T dwarfs with masses
of $\sim 1$~M$_{Jup}$. Traditional star formation theory suggests that
such objects should not exist, as noted above. However, they may exist
if formed in a different manner, eg. while experiencing strong 
photoevaporation from the O-type stars in the cluster centre (see investigations
by Whitworth \& Zinnecker 2004; Kroupa \& Bouvier 2003) or if they formed in a 
solar system as planets and were ejected following gravitational interaction with 
another planet or a passing star. 

In this survey we detect many new PMCs, including a candidate T dwarf.
The observations and the photometric results are described in sections
2 and 3. We also explore the implications for the IMF. Section 3 also contains a 
comparative spatial and binarity analysis for stars and brown dwarfs. In section 4
we report the discovery of two bipolar nebulae and discuss some of the trails 
and filaments of nebulosity which are observed in the survey area.
Our conclusions are given in section 5.

\section{Observations}

Near infrared imaging in the J (1.25~$\mu$m), H (1.65~$\mu$m) and K 
(2.2~$\mu$m) bands was carried out at 2 epochs with the Flamingos I camera 
(Elston 1998) on 
the Gemini South telescope at Cerro Pachon, Chile. All observations were made
in queue mode by observatory staff. The first dataset was taken on the nights
of 7, 10, 12, 15, 16 and 17 October 2001, at which time the Flamingos camera
employed an engineering grade 2048$^2$ HAWAII2 HgCdTe detector array.
One field was observed, the camera pixel scale of 0.076 arcsec yielding
a field of view of 156 arcsec. 
The second dataset, comprising 2 Flamingos fields, was taken in 2002 with a 
science grade array of the same type on the nights of 29, 30 and 31 October, 
and 1, 2, 4 and 12 November. A 2$\times$2 jitter pattern with $\sim 30$ 
arcsec offsets was used for all 3 fields. The location of the 3 contiguous fields, 
denoted Field 1, Field 2 and Field 3, is shown in Figure 1, superimposed
on a JCMT/SCUBA map of 850~$\mu$m emission from cold dust in the survey region (data
courtesy D.Johnstone, private comm., see Johnstone \& Bally 1999). The 
location of a less sensitive Gemini North/NIRI field, which we describe 
briefly in Section 3.2, is also indicated. Total exposure 
times for each Flamingos field were approximately 1 hour at K band, 1 hour 
at H band, and 2 hours at J band. Individual integration times, seeing data, 
and exact exposure times after discarding poor data are given in Table 1.
A 4th Flamingos field, offset by 30 arcminutes east, was requested in order to 
investigate the luminosity functon of contaminating background stars but 
these data were not taken due to contraints on availability of the instrument.

\begin{figure*}
\begin{center}
\begin{picture}(200,480)

\put(0,0){\includegraphics{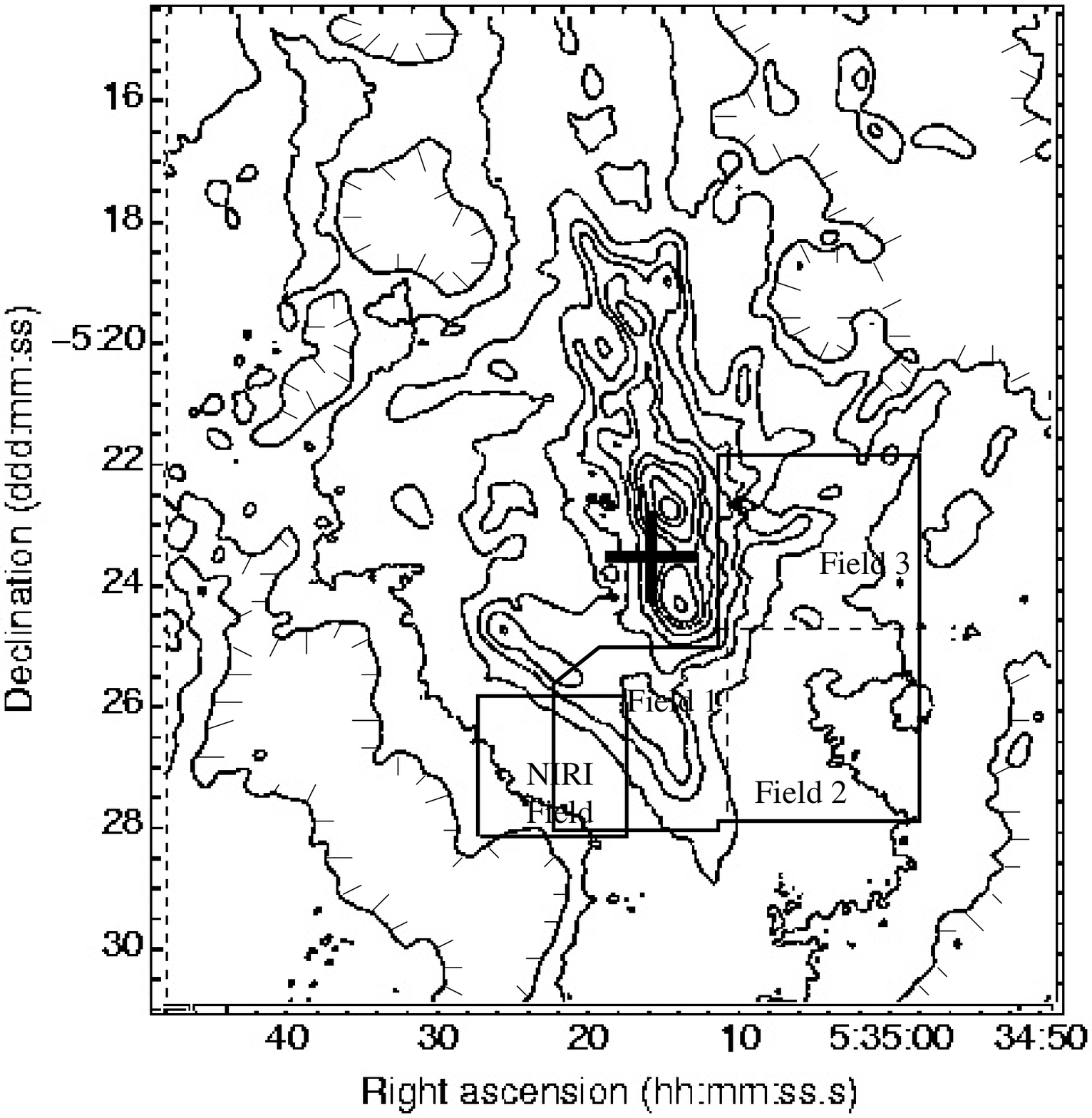}}

\end{picture}
\end{center}
Figure 1. Contour plot of the 850~$\mu$m emission from cold
dust in the OMC-1 region (see Johnstone \& Bally 1999.) The survey region is 
enclosed by solid lines and the locations of the 3 Flamingos fields are 
indicated (J2000.0 coordinates). Strong dust emission is expected to 
correlate with high extinction through the cloud, indicating regions where 
background stars with low reddening are less likely to be seen. The location
of $\theta_1$~Ori~C at the cluster centre is indicated by a large $+$ sign. The zero 
flux contour is marked with short lines. Contour levels (in data numbers) 
are 0, 0.2, 1, 2, 3, 5, 8, 12, 20, 50 and 100.  The 0.2 contour level
corresponds to visual extinction on the order of A$_V = 5$. Areas of negative 
flux are influenced by imperfect beam switching.
\end{figure*}

These fields were selected to avoid the bulk of the brightest nebulosity
in the Trapezium cluster in order to improve sensitivity and permit follow
up spectroscopy by reducing the surface brightness gradients in the 
background. The extinction through the backdrop of the OMC-1 cloud varies
greatly across survey area, as indicated by the 850~$\mu$m 
emission. Not all regions of high optical extinction are prominent in Figure 1: 
the ridge of high extinction seen by Hillenbrand (1997), running several arcminutes
southeast from (05h 16m, -5$^{\circ}$ 19m), appears merely as a narrow tongue 
extending from the main cloud. By contrast, the Orion Bar (located 3 arcminutes
southeast of $\theta_1$~Ori~C) appears 
prominent in Figure 1, suggesting it is a structural feature as well
as part of the ionisation front. Bright nebulosity in the bar reduced
sensitivity in the northern half of Field 1. 
The K band is most
sensitive to M and L type brown dwarfs in Orion, which have red colours
due to both their cool photospheres and reddening within the OMC-1
molecular cloud in many cases. The less dense medium
in front of the OMC-1 cloud, in which many of the cluster members lie,
produces much less extinction (O'Dell \& Yusef-Zadeh 2000). J and H band
data are necessary to determine the extinction toward each source, since
a significant fraction of the K band flux is produced by hot circumstellar
matter in many cases (Lada et al. 2004; Muench et al. 2001).

No photometric standards were observed except for the 1 or 2 per night taken 
for all Gemini programmes. Photometric zero points were calculated using the
large number of published surveys of the region instead (see below).
No detailed information on the photometric quality of the observing 
conditions was provided but the stability of the background count 
level in the individual integrations suggests that conditions were either
photometric or close to it during all the observations. All the Flamingos data 
were taken at low airmass, which aids photometric stability. However,
data taken in different filters for each field were generally taken
on different nights, so it is likely that T Tauri-like source variability 
will have affected the measured colours in a significant fraction of sources, 
mostly at the 0.1 to 0.2 mag level (see Section 3).

\subsection{Data Reduction}

The data were Dark subtracted and flat fielded using the IRAF software package
and areas of bad data were removed from each data frame with a mask. 
Approximately 92\% of the engineering grade array provided useable data.
In both the engineering and science grade arrays the first few dozen rows of 
each of the four 1024$^2$ array readout quadrants showed elevated counts due 
to amplifier glow. The amplifier glow was time variable, so a Dark frame with 
a similar amount of glow was assigned to each image for best subtraction. This 
minimised the amount of data that had to be masked out.

The K band data from the engineering grade array showed a large
spatially structured component of the background due to constructive 
interference by OH emission lines in the earth's atmosphere. This appeared
as a bright ring in the middle of each image, surrounded by much fainter
concentric rings in a Fabry-Perot interference pattern. These fainter rings
were visible only after flat fielding, if the structure were not first removed.
The interference pattern was sufficiently stable to permit a straightforward
subtraction by using data taken for photometric standards to make a
template interference pattern (the Orion data could not be used because of
the nebulosity). Where necessary the template was scaled by a factor close 
to unity to improve the quality of the subtraction. After subtraction of this 
feature the data were flatfielded with data taken in bright twilight, in which 
no sign of OH interference was apparent.

The K band data from the science grade array also showed a large spatially 
structured component of the background, but due to a ghost image of the
telescope secondary mirror and the mirror supporting spider. (This may also 
have been present at a lower level in the data from the 
engineering grade array). The unwanted signal was subtracted with a template
in the same way as the OH airglow pattern, before flatfielding with
twilight flats. The J and H band data from both arrays
showed only very low levels of the OH interference pattern, which
was undetectable amongst the nebulosity in most of observed area of
Orion. This was fortuitous since we did not have a suitable data to make 
a template for subtraction. Since the twilight flats did not contain a
detectable OH signal the small amount of OH interference structure remaining
in the final coadded data had a much smaller effect on the photometry of
faint sources than the pervasive nebulosity.

The corners of the arrays often showed an additional background glow, 
in the K band only, associated with the telescope guide star probe. This 
varied with the jitter position and affected different corners in the two 
epochs. Low level glow had little impact on the photometry since the flux
in a photometric aperture was easily subtracted with a concentric annulus.
However some small portions of the data had to be removed where the glow was
very strong, leading to gaps in the north west corners of the data mosaics for
Fields 2 and 3 at K band.

The data were coadded into a moasic for each field with the Starlink package 
CCDPACK. Point sources were detected by eye, since automated detection 
algorithms have difficulty in nebulous regions. The data contained a large
number of residual images from sources saturated at a previous jitter position
as well as numerous ghost images. Some ghost images were point source-like
but could be identified by their consistent locations relative to the
brightest stars in each field. Residual images also lay at predictable 
positions defined by the jitter pattern, and usually had a resolved disc-like 
appearance, confirmed by a very different radial profile from a point 
source with the IRAF utility IMEXAM. 

\subsection{Photometry}

Aperture photometry was then conducted with the APPHOT package
in IRAF. For the brighter sources a photometric aperture with a 13 pixel
radius was used, except in Field 3 where a 9 pixel aperture was used for
the K band data because of better seeing conditions. For these bright sources
an annulus at a radius of 45 to 50 pixels was employed for backgound 
subtraction, using the median of the background counts. For all fainter 
sources (including all those faint enough to be brown dwarf candidates) 
smaller photometric 
apertures were selected to optimise the signal to noise. Smaller annuli were 
used for background subtraction, in order 
to minimise the effect of the nebulosity gradients. The small sky annuli for 
faint sources often included a measurable fraction of the point source flux 
so this was carefully calibrated using several bright stars in each field, as 
were the aperture corrections. All sources were inspected individually
to ensure good photometry and detect any close binaries. The NSTAR routine
in IRAF was used for photometry of close binaries. 
In cases where one or more binary components were saturated the relative fluxes
of the two were estimated using the skirts of the image profiles, with 
the total flux being taken from previously published data.

The source of photometric zero point for each field is indicated in Table 1.
For each field we attempted to find several bright but unsaturated
sources with fluxes published in another survey. It is necessary to use several
sources because of the likelihood of variability in a significant fraction of 
objects. It was usually found that most sources in a sample produced a 
similar zero point (within 0.1 mag) but a minority were outliers, probably
due to variability, and were therefore discarded. 
The final formal uncertainty ($\sigma_{n-1}/n^{1/2}$) in the zero 
points for each filter in each field was less than 0.03 mag in each case,
though the comparison of different surveys suggests that the true uncertainty
may be larger (see below). No linearity data for the Flamingos detector arrays 
was supplied by the observatory, so a linearity correction could not be applied
during reduction. However, investigation of the zero points derived from
individual stars in each sample showed no trend as a function of well depth,
indicating that photometric errors arising from non-linearity are unlikely to 
exceed 2\% for sources below 80\% full well. This is as expected for 
HgCdTe HAWAII arrays.
The 2MASS catalogues were the default choice for zero point calculations  
due to the complete spatial coverage of the 2MASS survey. Both the
2MASS 2nd Incremental Release and the All Sky Survey were searched for 
counterparts to our sources (since the latter was published during this
analysis), using the profile-fitted 2MASS fluxes. The colour transformations
of Carpenter (2003) were used to correct the zero points from the 2MASS system
to the MKO system, assuming (J-K) colours close to unity for the photospheres 
of the sources in each sample, as indicated by the NextGen isochrone colours
shown in section 3.
In the J and H bands we found that the two 2MASS surveys provided 
internally consistent zero points (based on 7 to 10 sources per field) which
agreed well with each other. Fluxes from the 2nd Incremantal Release 
sometimes had less scatter in the zero points derived from each star than the 
All Sky Survey (which has fluxes based on a reprocessing of the same data) but 
this was counterbalanced by the larger number of sources extracted in the 
Orion region in the All Sky Survey.

At K band neither 2MASS catalogue produced an internally consistent zero point 
for Fields 1 and 3.
Field 2 contains less nebulosity than the other fields. Brighter nebulosity 
may explain why the 2MASS fluxes showed poor consistency in the other 2 fields, 
especially since the brightest sources (for which nebulosity has little effect on 
photometry) were often saturated in our data. For Field 1 we 
therefore used the dataset of Hillenbrand \& Carpenter (2000) to provide the K 
band zero point. We found that a sample of 14 sources in the northern third of 
the field (the region of overlap with our data) yielded a very consistent zero 
point with formal uncertainty of 0.013 mag, after excluding 4/14 photometric 
outliers from the calculation. For Field 3, the longer integration time per 
data frame and excellent spatial resolution meant that more bright sources 
were saturated so fewer sources were available for photmetric comparison.
Comparison with the overlapping region of the Hillenbrand \& Carpenter dataset
showed no internal consistency for sources with 
K magnitudes $13<$K$<16$. Since observing conditions were flagged by the 
observers as photometric we therefore used the single Gemini photometric standard to 
define the zero point, and checked it against the brightest sources in the 
Hillenbrand \& Carpenter dataset that were unsaturated in our data.
Agreement to better than 0.02 mag was found 
for the 2 sources in that dataset with $13<$K$<14$. An additional check for 
all 3 fields at all 3 wavelengths was made by comparison with the JHK 
photometry of Muench et al.(2002, NTT dataset). The K band zero point 
for Field 3 derived from 18 sources with $13<$K$<16$ agreed to within 0.012 
mag with our adopted value. Note that the K band filter used in the  
Hillenbrand \& Carpenter (2000) Keck data was judged sufficiently similar to 
the MKO K band that no photometric transformation was necessary.

Comparison of our 9 zero points for the 3 fields with those
derived from the Muench et al.(2002) data showed that 7/9 agreed
to within 0.05 mag but the H and K zero points for Field 2 differed
by approximately 0.1 mag. Similarly, comparison with the J and H band data of 
LR00 and Lucas et al.(2001), which were taken on several different nights, 
yielded close agreement for 5/6 zero points but a 0.1 mag disagreement
for Field 3 in the J band.
We therefore caution that the fluxes presented in this paper may have
errors of $\sim 0.1$ mag arising from the zero points, though the 
similarity of the (J-H) vs. (H-K) two colour diagrams for the 3 fields suggests 
that the errors are probably less than this. With regard to source
colours and dereddening analysis, variability is likely to have produced
larger errors ($\sim 0.1$ mag or more) than the process of photometry in 
perhaps half of all sources, see Carpenter, Hillenbrand \& Skrutskie 2001;
Kaas 1999).

\section{Results}
\subsection{Colour-magnitude Diagrams}

\begin{figure*}
\begin{center}
\begin{picture}(200,250)

\put(0,0){\includegraphics{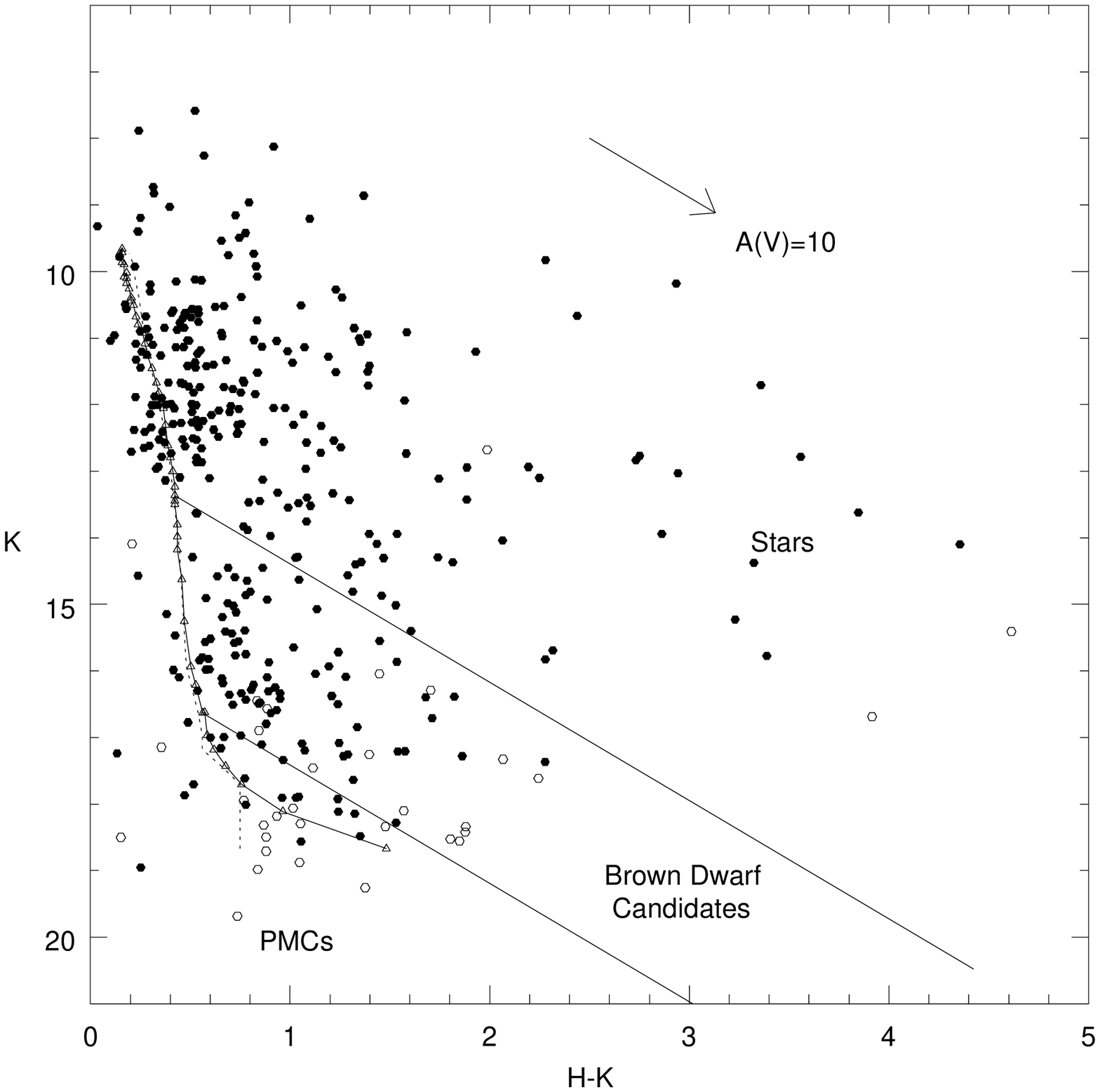}}

\put(0,0){\includegraphics{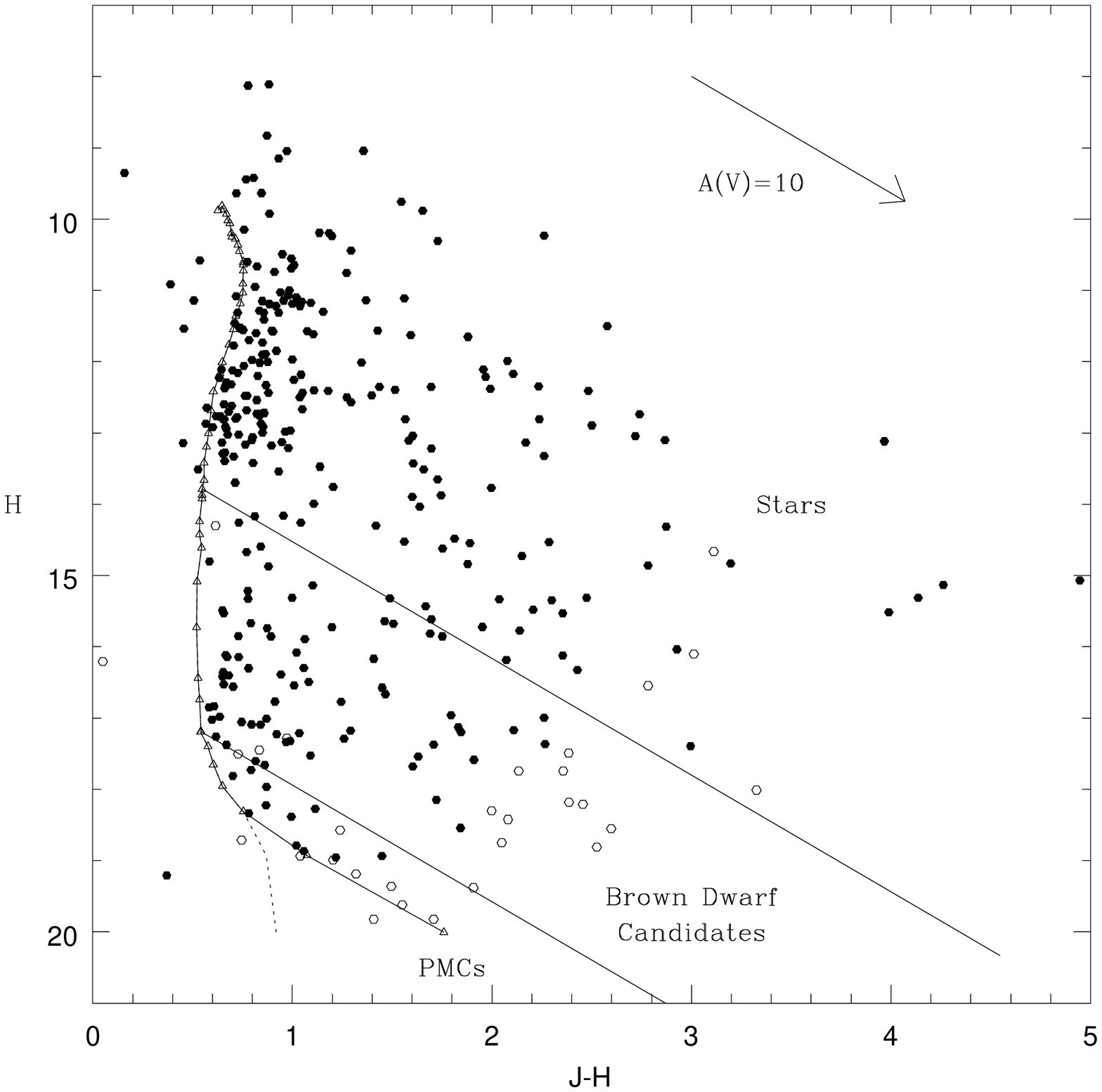}}

\end{picture}
\end{center}
{\it (left)} Figure 2. {\it (right)} Figure 3. The colour magnitude digrams
show all 396 sources detected in the survey region (filled circles for
good data and open circles for data with uncertainties $>0.2$ mag along either 
axis). The solid line connecting the triangles shows a combined NextGen 
and AMES-Dusty theoretical 1~Myr isochrone (see text) while the dashed lines 
show the colours of mature field dwarfs with the same temperatures as a check 
on the calculation. The reddening vector in each figure shows the effect of 10 
magnitudes of visual extinction on the data points. A wide range of extinctions is 
apparent, from A$_V = 0$ up to A$_V = 50$. Solid lines parallel to
the reddening vector are used to separate stars, brown dwarf candidates
and planetary mass candidates.

\end{figure*}

In Figures 2 and 3 we present the photometric results as K vs.(H-K) and 
H vs.(J-H) colour magnitude diagrams. Data points are shown by filled circles,
or open circles for the most uncertain points. For saturated sources we use
data from the 2MASS archive and the surveys of Hillenbrand \& Carpenter (2000)
and Muench et al.(2002). Fluxes and coordinates are given in Table 2\footnote{
The full Table 2 is available in the electronic on-line version. H band images
located at http://star-www.herts.ac.uk/$^{\sim}$pwl/gemini.html can be used to locate 
sources.}. A theoretical 1~Myr isochrone 
for masses ranging from 1.2~M$_{\odot}$ to 0.003~M$_{\odot}$ (3~M$_{Jup}$) is
plotted as a solid curve at the left of each colour magnitude diagram and solid 
lines parallel to the reddening vector are used to distinguish between stars, brown 
dwarf candidates (M$<0.075$M$_{\odot}$) and planetary mass candidates (PMCs)
(M$<0.012$M$_{\odot}$). Substellar sources are referred to only as candidates 
since the great majority have not yet been observed spectroscopically 
and there is likely to be a significant proportion of background stars among
the faintest sources in this very deep survey (see section 3.2). 
396 point sources were detected in total of which 390 were seen at K band, 365
at H band and 326 at J band. The presence of $\sim 33$ PMCs is apparent 
(mostly new detections) and we confirm the detection all sources previously 
published within the area of this survey in LR00 and Lucas et al.(2001). 

To construct the theoretical isochrone we used a combination of the NextGen 
isochrone (Baraffe et al.1998; Chabrier et al.2000) at effective temperatures T$>2200$~K 
and the DUSTY isochrone of (Chabrier et al.2000)) at  T$<2200$~K. The dust 
free NextGen atmospheres provides a better fit to the colours of mature field 
dwarfs above 2200 K but the DUSTY atmospheres are better at 1800 to 2200~K,
corresponding to 4.5 to 9~M$_{Jup}$ on the 1~Myr isochrone. 
(Note that we have added a correction of 0.15 mag to the
(J-H) colours of the DUSTY isochrone. This was indicated for field
dwarfs (Chabrier et al.2000) and permits a smooth joining with the NextGen isochrone
in Figure 3.) Below 1800~K the DUSTY model predictions become much redder than 
is observed in field dwarfs. The same effect is seen here where the isochrone curves
through the faintest data points at K$>16.5$ in the K vs.(H-K) diagram.
This is unsurprising since the DUSTY model represents the limiting 
case of high dust opacity. In the K vs.(H-K) diagram we overplot the
observed colours of field stars and brown dwarfs as a dashed line, using data
from Knapp et al.(2004); Leggett et al.(2001), Dahn et al.(2002) and Tokunaga (2000).
The colour transformations of Carpenter (2002) were used to convert to the MKO
photometric system.
It is clear that the much lower surface gravity of 1~Myr sources has little 
effect on the (H-K) colours at T$>1800$~K, so we can adopt the colours of field
L dwarfs with some confidence for cooler 1~Myr old objects with masses in the range
3-5~M$_{Jup}$, for which the isochrone predicts $1900>$T$>1550$~K.
Therefore when calculating dereddened fluxes we deredden to the theoretical
isochrone at T$>1900$~K (solid lines) but adopt field dwarf colours 
(dashed lines) for T$<1900$~K, corresponding to M$<5$~M$_{Jup}$.

The theoretical isochrone in the H vs.(J-H) diagram lies to the left
of almost all the data points. The NextGen isochrone may be slightly too blue at 
brown dwarf masses (see section 3.4) since there is a slight gap ($\sim 0.1$ mag) 
between the isochrone and the data points at $13<H<17$. Main sequence colours are 
again adopted for the coolest sources with T$<1900$~K, as shown by the dashed extension 
to the theoretical isochrone. Two faint sources show very blue colours in the H vs.(J-H) 
diagram, lying well to the left of the isochrone. One of these is an uncertain data 
point shown by an open circle at 
H=16.2, J-H=0.05; it lies atop a pedestal of nebulosity that undoubtedly 
contaminates the point source flux. The other, Source 318, has H=19.21, J-H=0.37, 
H-K=0.25, and is a candidate T dwarf. Source 318 is discussed further in section 
3.5. A few much brighter sources also lie to the left of the isochrone in Figure 3 
but these data points are mostly drawn from other surveys (being saturated 
in our images) so it is difficult to comment upon them. A significant number of 
data points derived from this survey lie 0.2-0.3 mag to the left of the 
isochrone in Figure 2. Some of these have uncertain photometry 
(open circles): some lie atop highly structured nebulosity and some appear 
slightly resolved at 1 wavelength, suggesting that they are very 
young sources embedded in circumstellar matter. Other sources with apparently 
well measured fluxes are likely to be due to either: (a) photometric scatter
from the wings of the error distribution in this large dataset 
(particularly at K band where the background subtraction was difficult); or
(b) the intrinsic spread in the colours of the photospheres, which is 
approximately 0.3 mag in field L dwarfs; or (c) variability, since fluxes in 
the different filters were often obtained on different nights.

It is preferable to deredden the observed fluxes using the H vs.(J-H) diagram
since the (H-K) colours of some faint sources probably have excesses 
attributable to hot circumstellar matter (see section 3.4). However the 
K vs.(H-K) diagram must be used for sources which are not detected at J band, 
and we also use it for sources which have very low signal to noise at J band 
but good H and K band photometry. The mean (H-K) colour excess for the brown
dwarfs in our sample is close to zero (see section 3.4) so we do not attempt to
correct for it when dereddening sources detected in only 2 colours. Note that 
PMCs with K band excesses may lie to the right of the dividing line between 
brown dwarfs and PMCs in Figure 3 and therefore appear overmassive.

\subsection{Luminosity Functions and Background Contamination}

\begin{figure*}
\begin{center}
\begin{picture}(200,220)

\put(0,0){\includegraphics{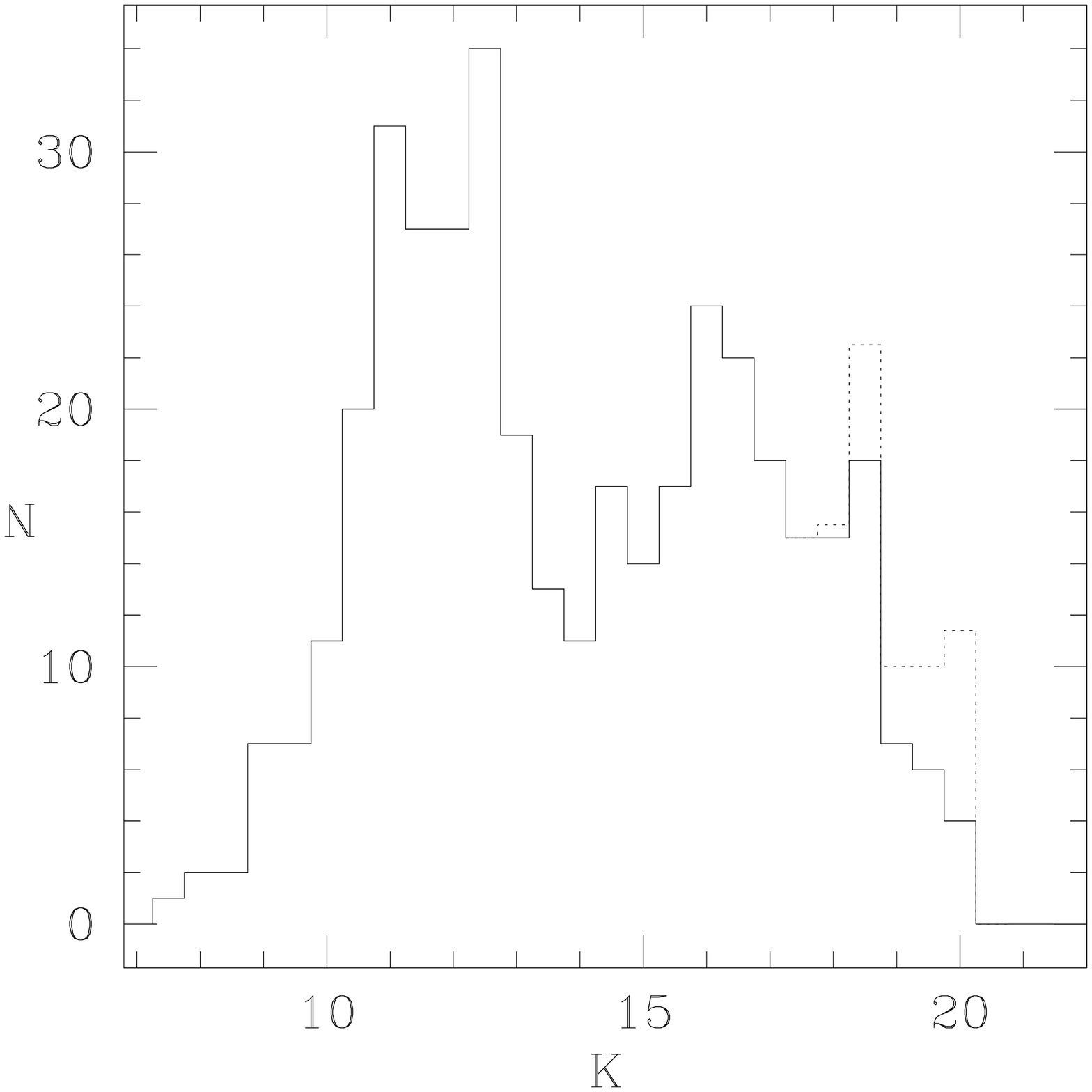}}

\put(0,0){\includegraphics{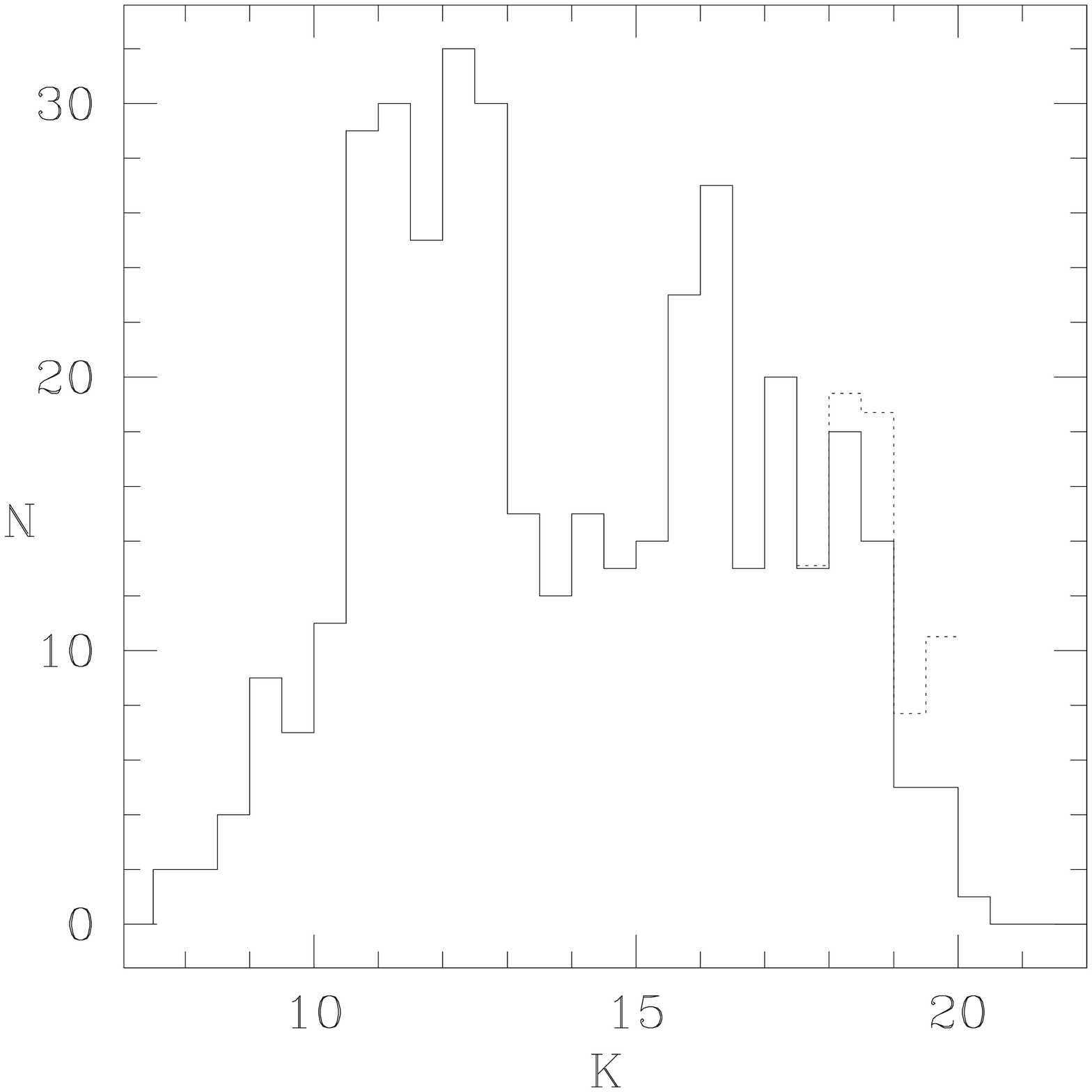}}

\end{picture}
\end{center}
Figure 4. Observed K band Luminosity Function. The histogram is shifted
by half a bin between the 2 plots to illustrate the shot noise. 
The solid line histogram shows the data while the dashed histogram
shows the effect of correcting for incompleteness at faint magnitudes.
\end{figure*}

\begin{figure*}
\begin{center}
\begin{picture}(200,220)

\put(0,0){\includegraphics{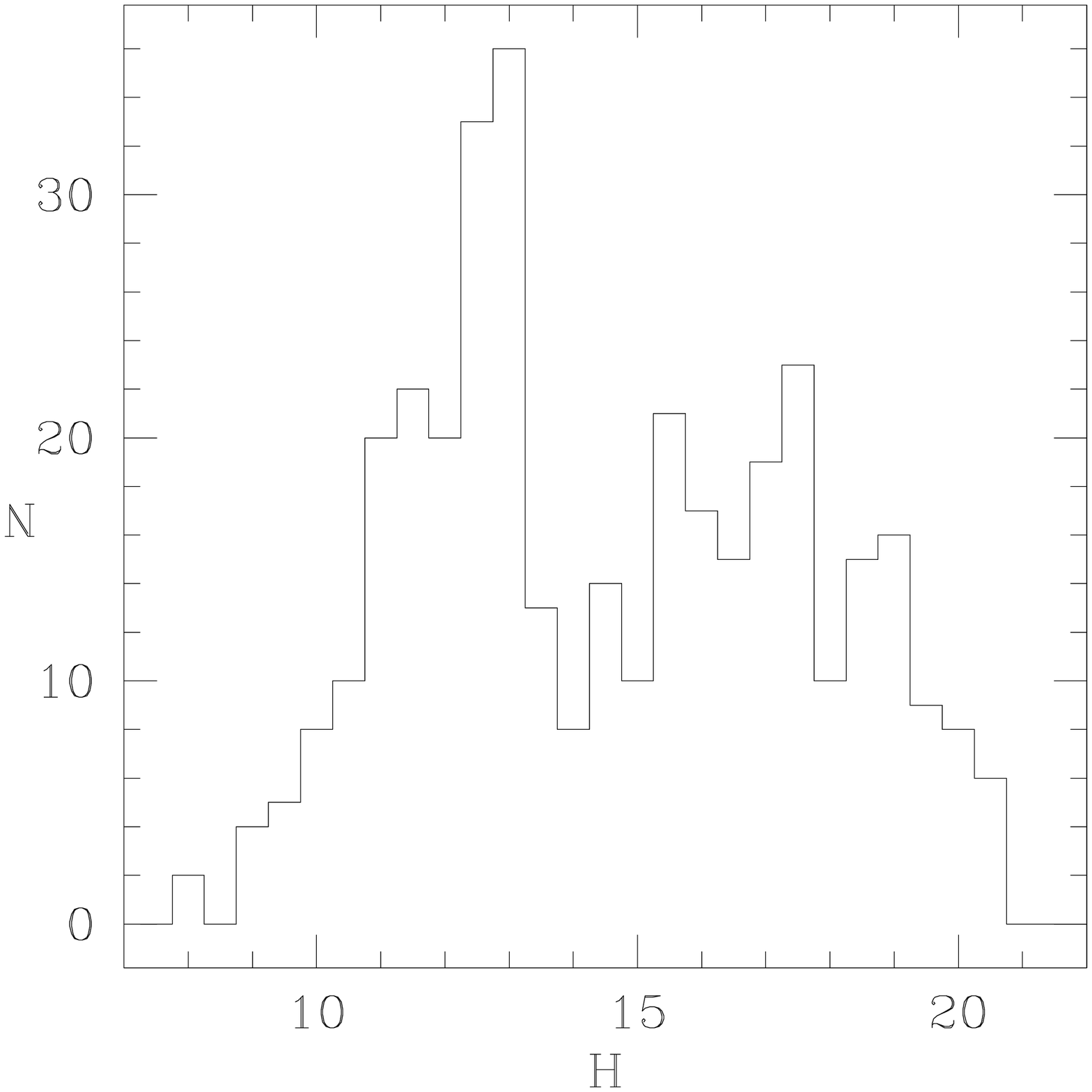}}

\put(0,0){\includegraphics{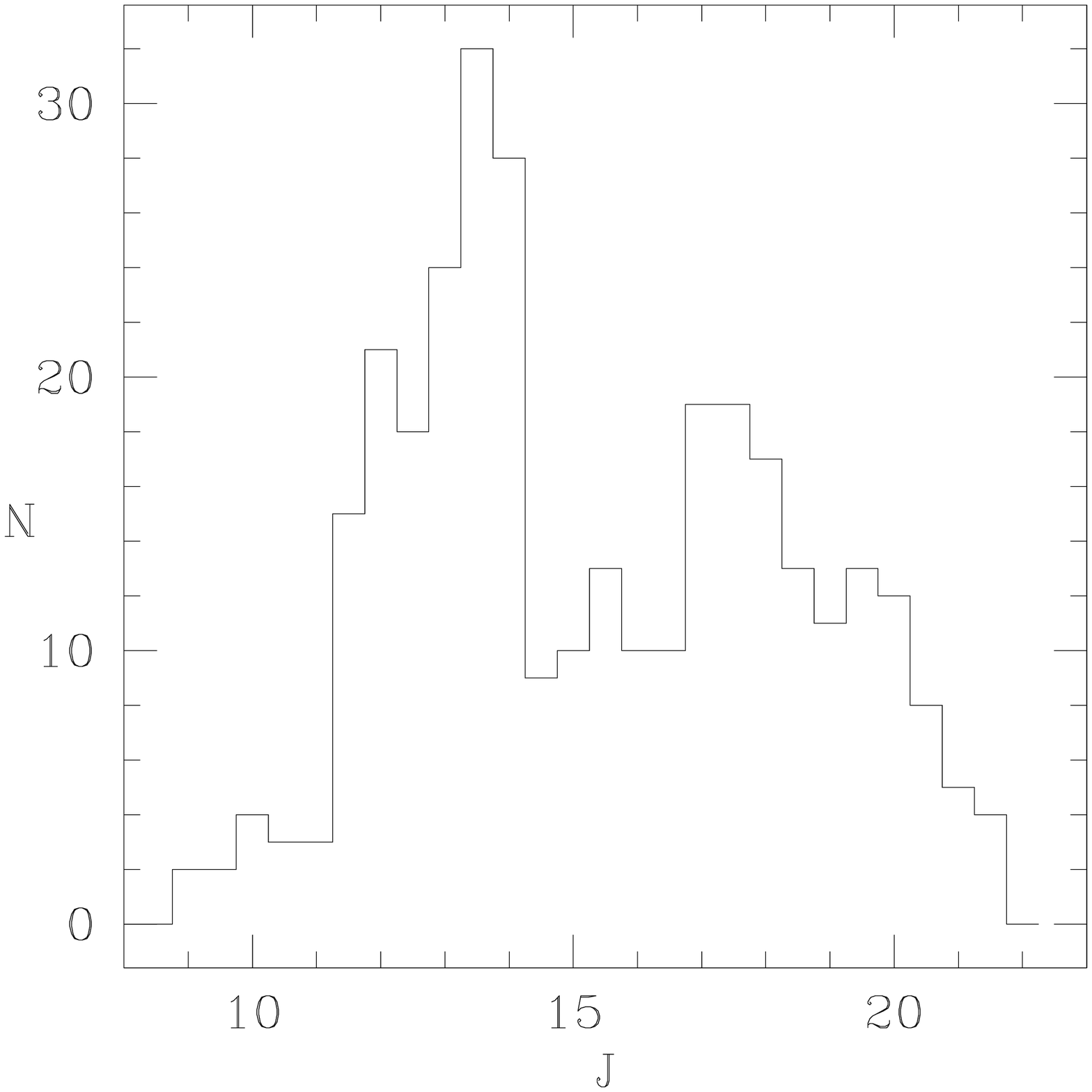}}

\end{picture}
\end{center}
Figure 5. (left) Observed H band Luminosity Function. 
(right) Observed J band Luminosity Function. 
\end{figure*}

In Figure 4 we plot the K band luminosity function, using two panels to show
the effect of shifting the bins by half a bin width. The H and J band
luminosity functions are plotted in Figure 5. The solid histogram of Figure 4
shows the actual data, while the dashed histogram at the faint magnitudes
shows the inferred luminosity function after correcting for incompleteness 
(see below). The stellar peak at 
K=11-12 and the brown dwarf peak at K=16 are apparent. These features have 
been previously reported by several authors for samples dominated by the 
central regions of the Trapezium cluster at r$<2.5$ arcminutes from 
$\theta_1$ Ori C (eg. McCaughrean et al.1995; LR00, Muench et al. 2002). It 
is interesting to see a similar luminosity function in this survey of the 
outer cluster (r=2 to 5 arcmin). This result is not very surprising since the
extreme youth of the cluster has allowed little time for dynamical mass
segregation. The cluster is only a few crossing times old for an 
assumed age of a few Myr and significant dynamical mass segregation
is not expected until an age $>10$~Myr (eg. de Grijs et al. 2002a; 2002b; 2002c).
There is a hint of a third peak near K=18.5, which appears independent of the 
binning after correcting for incompleteness. However this possible third peak
does not rise very significantly above the level of shot noise so it is 
unlikely to be real.

\begin{figure*}
\begin{center}
\begin{picture}(200,240)

\put(0,0){\includegraphics{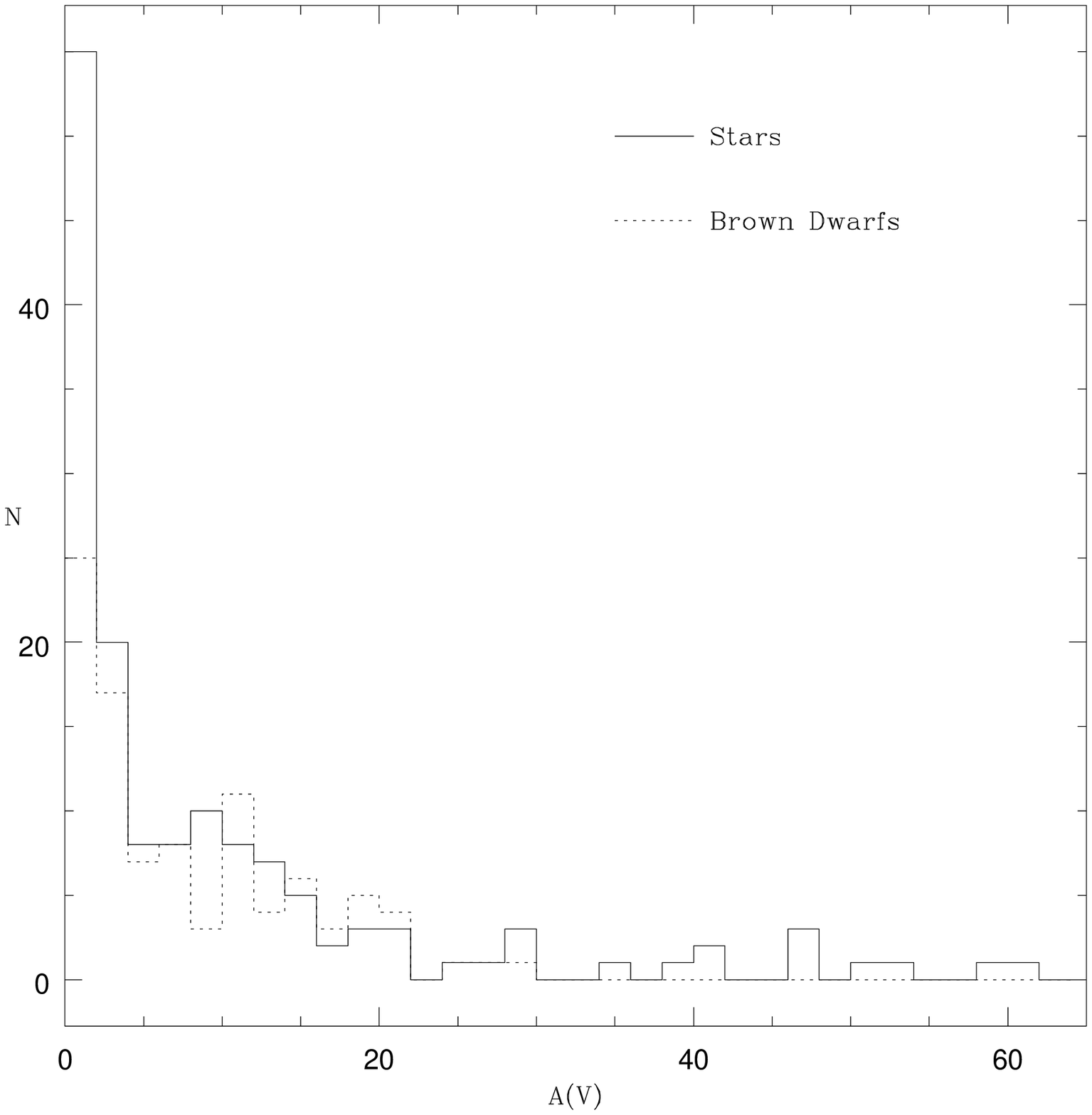}}

\end{picture}
\end{center}
Figure 6. Histograms of the measured extinction towards the stars (solid line)
and brown dwarfs (dashed line). PMCs are excluded since all have low extinction.
The brown dwarf histogram is flatter than the stellar histogram at 
$12< A_V <20$, despite rising incompleteness at $A_V>12$. This suggests
that the brown dwarf sample includes some background stars at $A_V>10$.  
\end{figure*}

\begin{figure*}
\begin{center}
\begin{picture}(200,270)

\put(0,0){\includegraphics{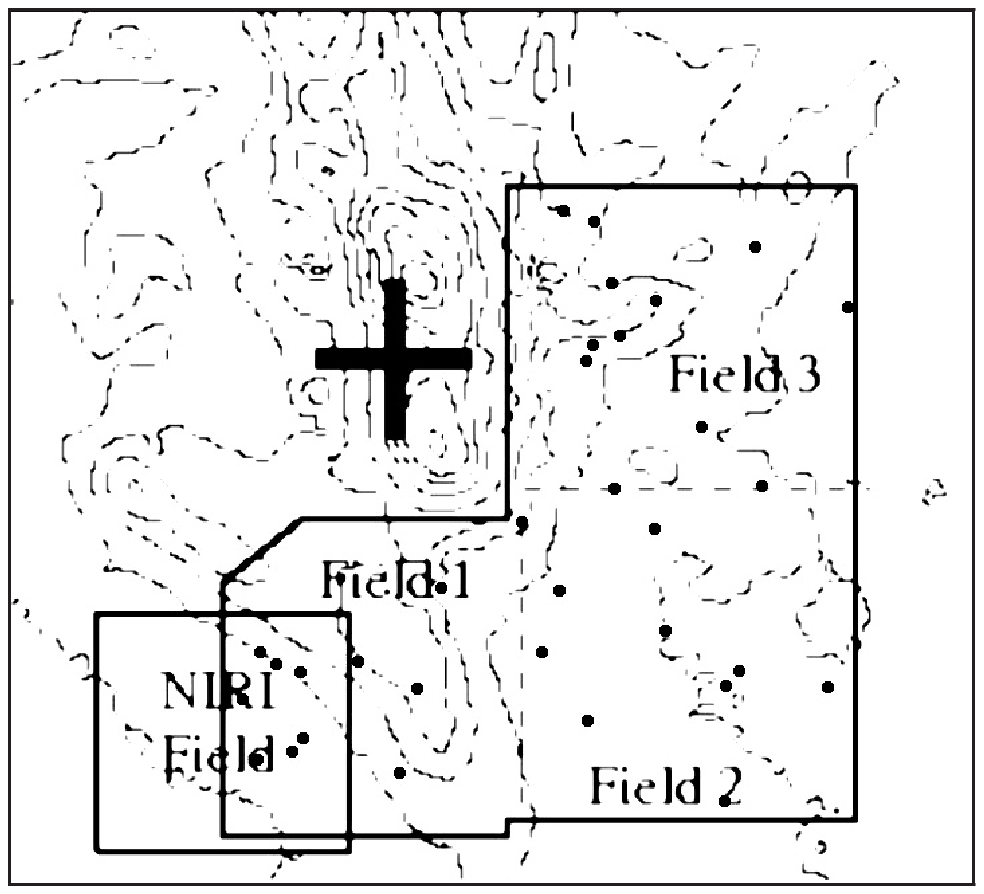}}

\end{picture}
\end{center}
Figure 7. Magnified view of Figure 1, with the location of the 33 PMCs shown
as filled circles. PMCs are difficult to detect in the northern half of
Field 1 due to the bright Orion Bar. Sources can be more clearly located
using the on line data at http://star-www.herts.ac.uk/$^{\sim}$pwl/gemini.html.
\end{figure*}

The analyses of Muench et al.(2002, based on photometry of an offset field) 
and Hillenbrand \& Carpenter (2000, based on a model of the galactic stellar 
population) both indicate that contamination by background objects starts to
become significant at about K=16 (or perhaps a little fainter, depending
on the assumed extinction through the backdrop of OMC-1) and then rises
to fainter magnitudes, perhaps peaking at K=19 to 20 (Hillenbrand \& 
Carpenter 2000). Given that the observed luminosity function declines
beyond the brown dwarf peak at K=16 despite the presence of background 
contamination it is probable that the cluster luminosity function declines 
steeply.

When considering the luminosity function in the planetary mass regime it
is necessary to take account of incompleteness. Completeness was
calculated by generating large numbers of artificial stars with the ADDSTAR
routine in IRAF and then inspecting the images to see whether the sources
were clearly detected. Only sources detected at $>5$$-\sigma$ in at least 1
filter (defined using the local background noise) are included in the catalogue 
in Table 2. Photometry 
at $<5$$-\sigma$ in the J or H bands is included if there is a $>5$$-\sigma$ 
K band detection. However, many possible K band detections with 
fluxes much greater than the $>5$$-\sigma$ detection threshold were excluded 
because the presence of highly structured nebulosity in most of the survey 
region made it difficult to distinguish low signal to noise detections from 
local peaks in surface brightness. This was a somewhat subjective judgement,
in which we were careful to exclude detections which might be false from the 
catalogue. This subjective element has little significance for the derived
IMF since detection in at least 2 filters is required to calculate a dereddened
luminosity and almost all the faintest H band detections were clearly
seen at K band. When assessing completeness with artificial stars we compared the data
with sources which were included in the catalogue to ensure consistency.
The experiments with ADDSTAR showed that completeness is patchy, with 
Field 1 being approximately 1 magnitude less sensitive than the other 2 
fields owing to relatively bright nebulosity and the poorer quality of data 
from the engineeering grade array. Completeness averaged over the whole survey
region declines from 80\% at K=18.5 to 70\% at K=19.0, 60\% at K=19.5
and 35\% at K=20.0. (These figures have a precision of approximately 5\%,
eg. completeness is $80\% \pm 4\%$ at K=18.5.)

The solid line histograms shows a marked decline in the luminosity 
function at K$\ge19$, which after correction for incompleteness (dashed lines)
appears rather as a flattening off at a level 
of $\sim 10$ sources per bin at K$\ge19$. Since contamination by background
stars is expected to introduce a relatively large number of stars at these
magnitudes it seems likely that the luminosity function for cluster members
declines toward fainter magnitudes in the planetary mass regime. There is no 
evidence for a very large population of planetary
mass objects such as that suggested by Bejar et al.(2001) for the 
$\sigma-$Orionis cluster. 

Unfortunately the uncertainties in the extinction through the
OMC-1 cloud on different sight lines through the survey region have the consequence
that even if very sensitive data 
were available for offset fields it would be impossible to calculate the 
number of background stars with high confidence. Histograms of measured
extinction are plotted in Figure 6 for stars and brown dwarfs (excluding PMCs)
to provide some insight into the extinction in our survey region. The solid line
histogram for stars (which is expected to include hardly any background stars) 
declines smoothly from $A_V$ = 10 to $A_V$ = 20. If there were no background stars
among the brown dwarf candidates then we might expect a slightly steeper
decline in the dashed histogram for brown dwarfs in this range, since
the sample begins to suffer from incompleteness at $A_V > 12$. However, the 
dashed histogram remains fairly flat in the $A_V$ = 12 to 20 range, suggesting
that there is a significant number of background stars with $A_V > 10$ 
(and K mag $>16$) among the brown dwarf candidates. Alternatively, the relatively 
high proportion of brown dwarf candidates with $10< A_V <20$ could 
conceivably be due to circumstellar matter, since it has been suggested that a 
significant fraction of proto-brown dwarfs are obscured by their own highly flared 
accretion discs (Walker et al.2004). In any event, the probable existence of
a background population with $A_V > 10$ provides little information on the
contamination among PMCs, since the sensitivity limits cause almost all of 
these have $A_V < 10$. The detection of a significant fraction of brown dwarf
candidates and stars with $A_V >10$ indicates that the OMC-1 cloud provides
an obscuring screen which removes background stars from the PMC
sample over much of the survey area. However, some of this extinction may be 
intrinsic to very young protostars and proto-brown dwarfs rather than due to the
screen. Moreover, observations of star formation regions much closer to the
galactic plane than the Trapezium cluster (b=-20$^{\circ}$) such as Serpens 
(Kaas 1999) and Chamaeleon I (unpublished observation by these authors) have 
shown that the extinction through a cloud can be very patchy on spatial scales 
less than 1 arcminute.

The locations of the PMCs are indicated as filled circles in Figure 7. The 
distribution of PMCs is fairly uniform, though there is some sign of an excess 
of PMCs toward the eastern edge of Field 1. This might have been
interpreted as a sign of a general increase in background contamination
with increasing distance from the dense ridge of OMC-1. However, visual 
inspection of the partially overlapping field observed with Gemini North/NIRI 
showed that this group of faint sources does not extend further to the east. 
Therefore it is likely just a chance grouping.

Since background contamination is difficult to calculate reliably, 
spectroscopic follow up of a large sample of PMCs is essential to determine 
the luminosity function of cluster members. We are presently in the process
of obtaining such data. It is possible to reduce the effect of background 
contamination by using a subsample with low reddening but even this method
is imperfect, as discussed in the next section.

\subsection{The Initial Mass Function}
\begin{figure*}
\begin{center}
\begin{picture}(200,445)

\put(0,0){\includegraphics{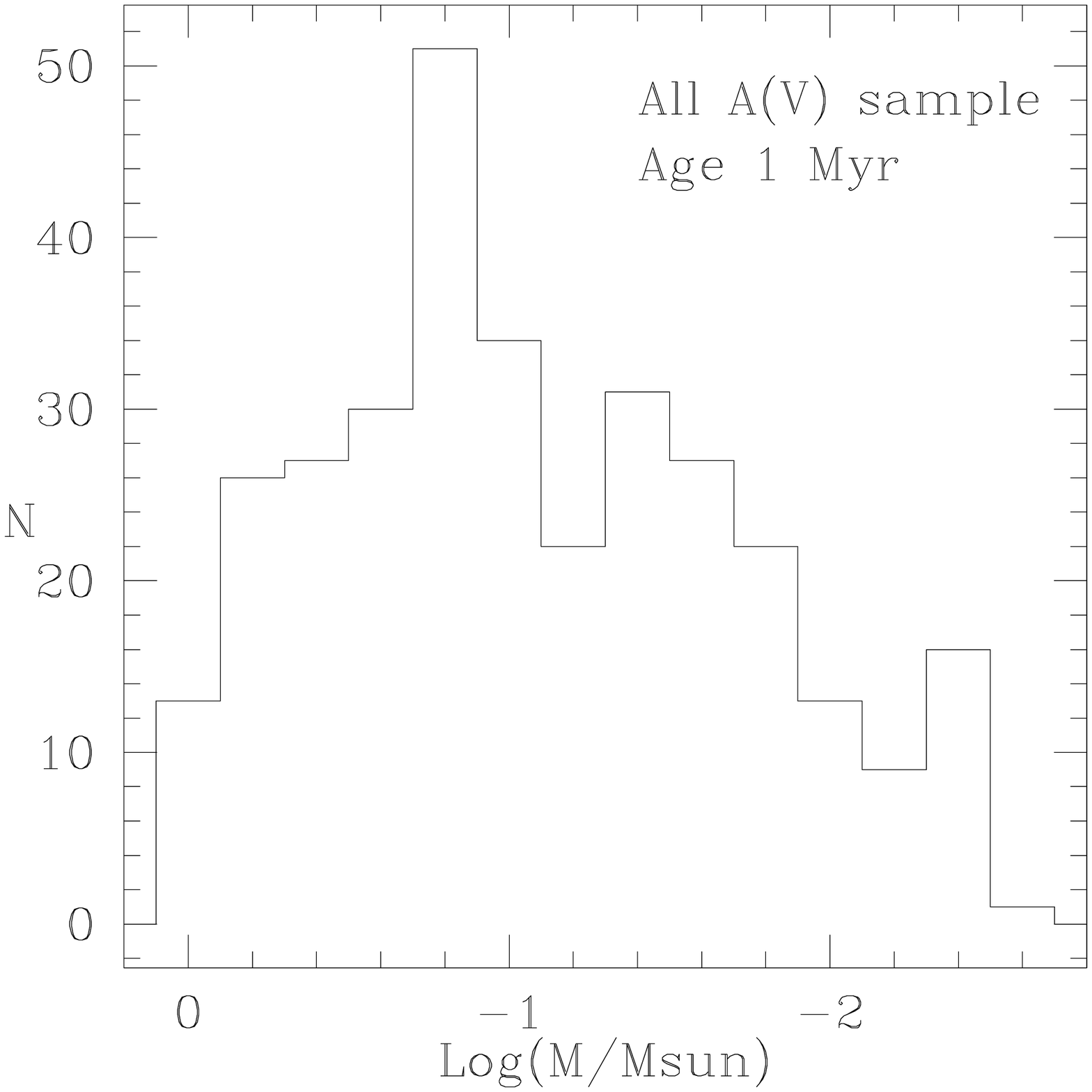}}

\put(0,0){\includegraphics{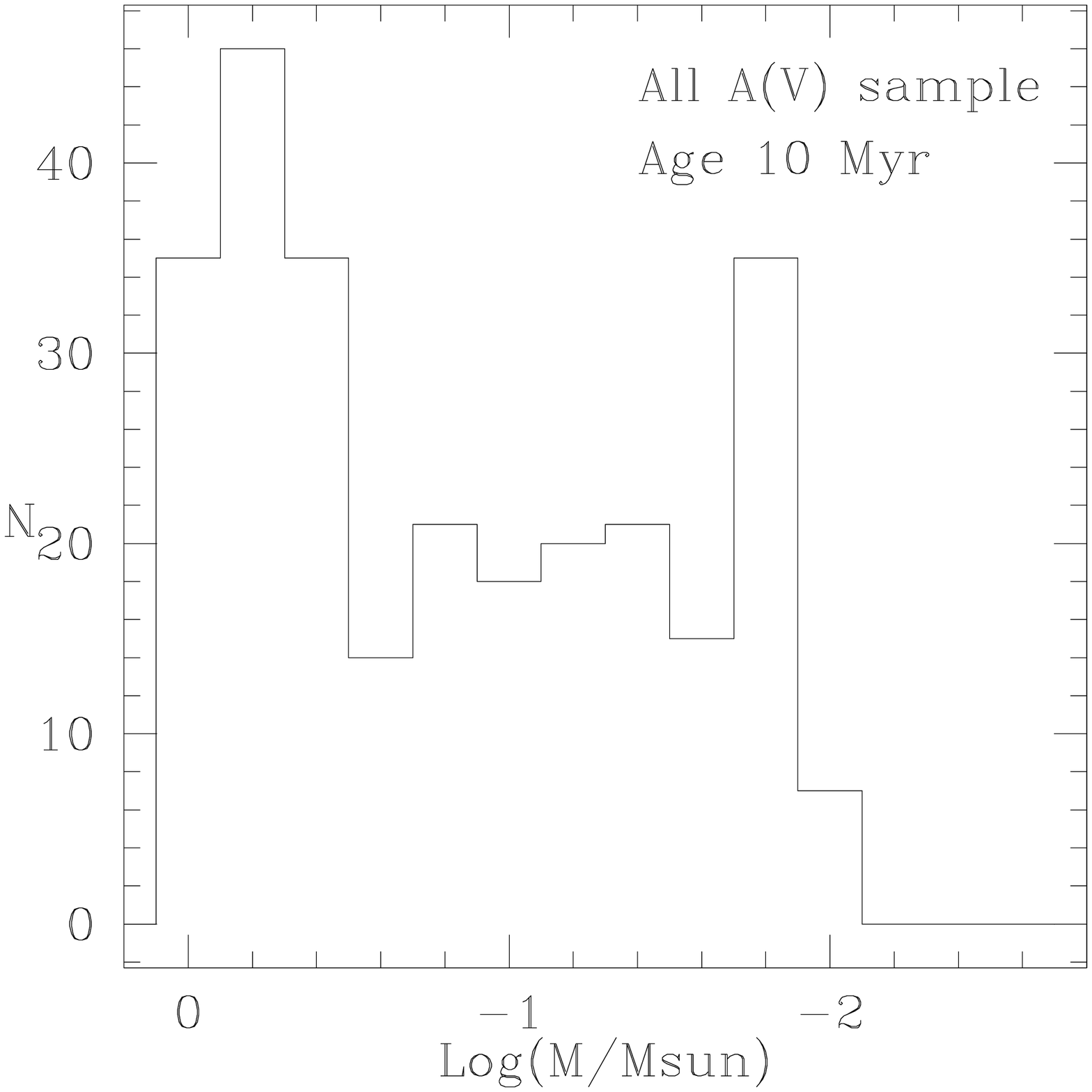}}

\put(0,0){\includegraphics{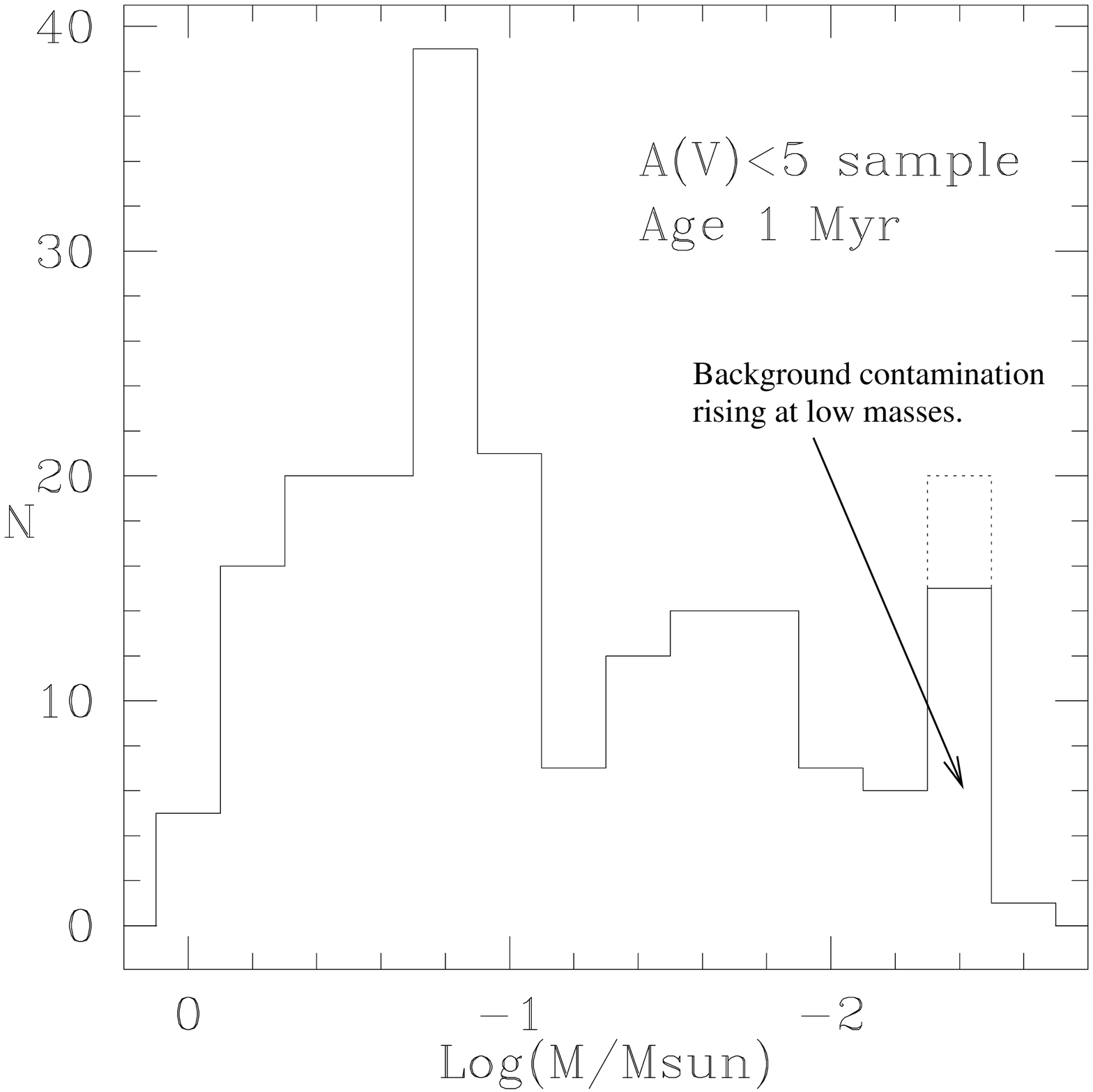}}

\put(0,0){\includegraphics{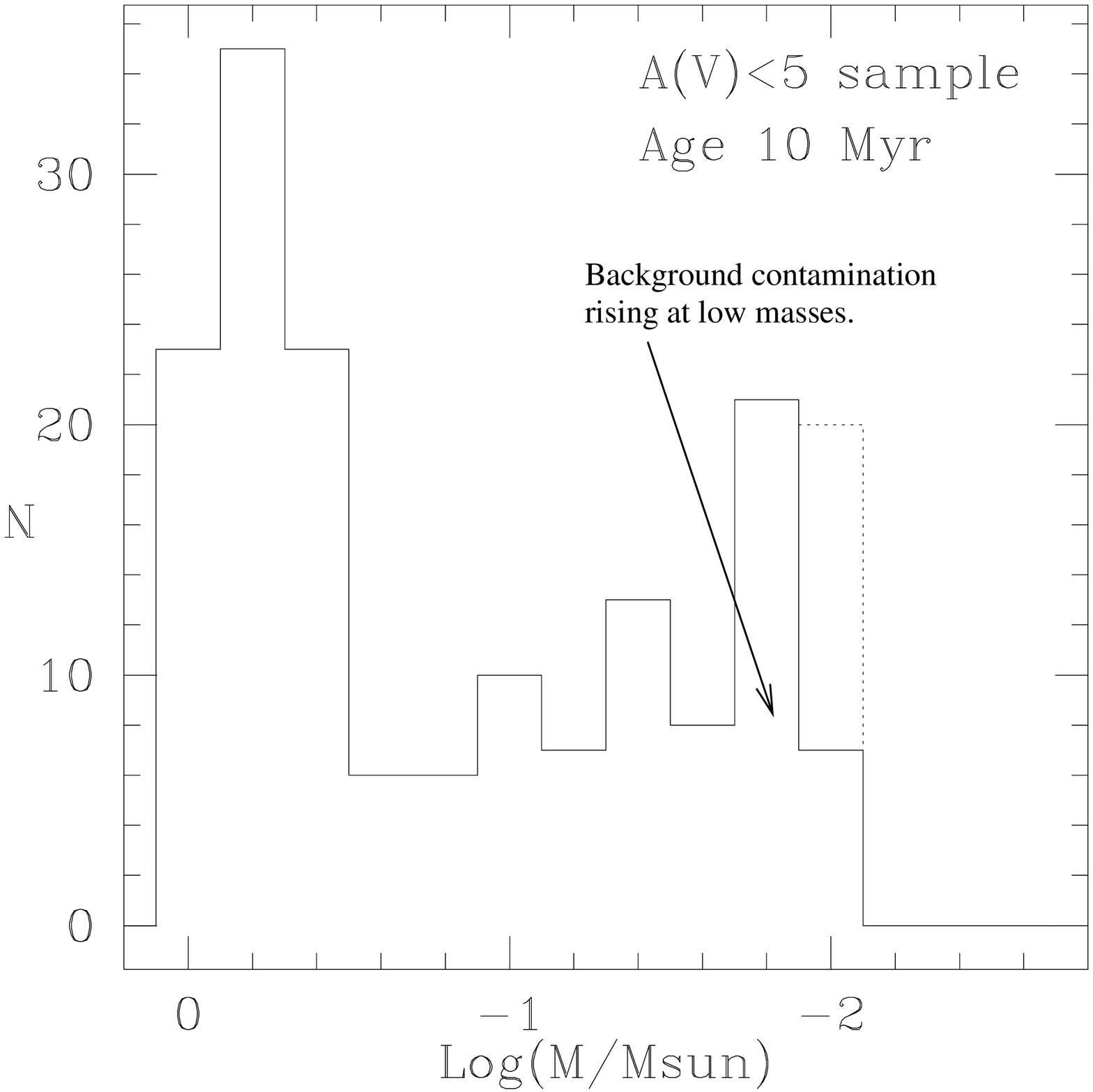}}

\end{picture}
\end{center}
Figure 8. Derived mass functions for different assumed ages. 
(left) 1 Myr. (right) 10 Myr. Upper panels show all sources detected. Lower panels 
show subsamples with A$_V < 5$, to reduce background contamination. For the 
A$_V < 5$ subsamples it is possible to correct for incompleteness at the very low mass 
end: the effect is shown by the dashed histogram in the lower panels. The apparent 
rise in the mass functions at very low masses is attributed to background 
contamination.
\end{figure*}

\begin{figure*}
\begin{center}
\begin{picture}(200,220)

\put(0,0){\includegraphics{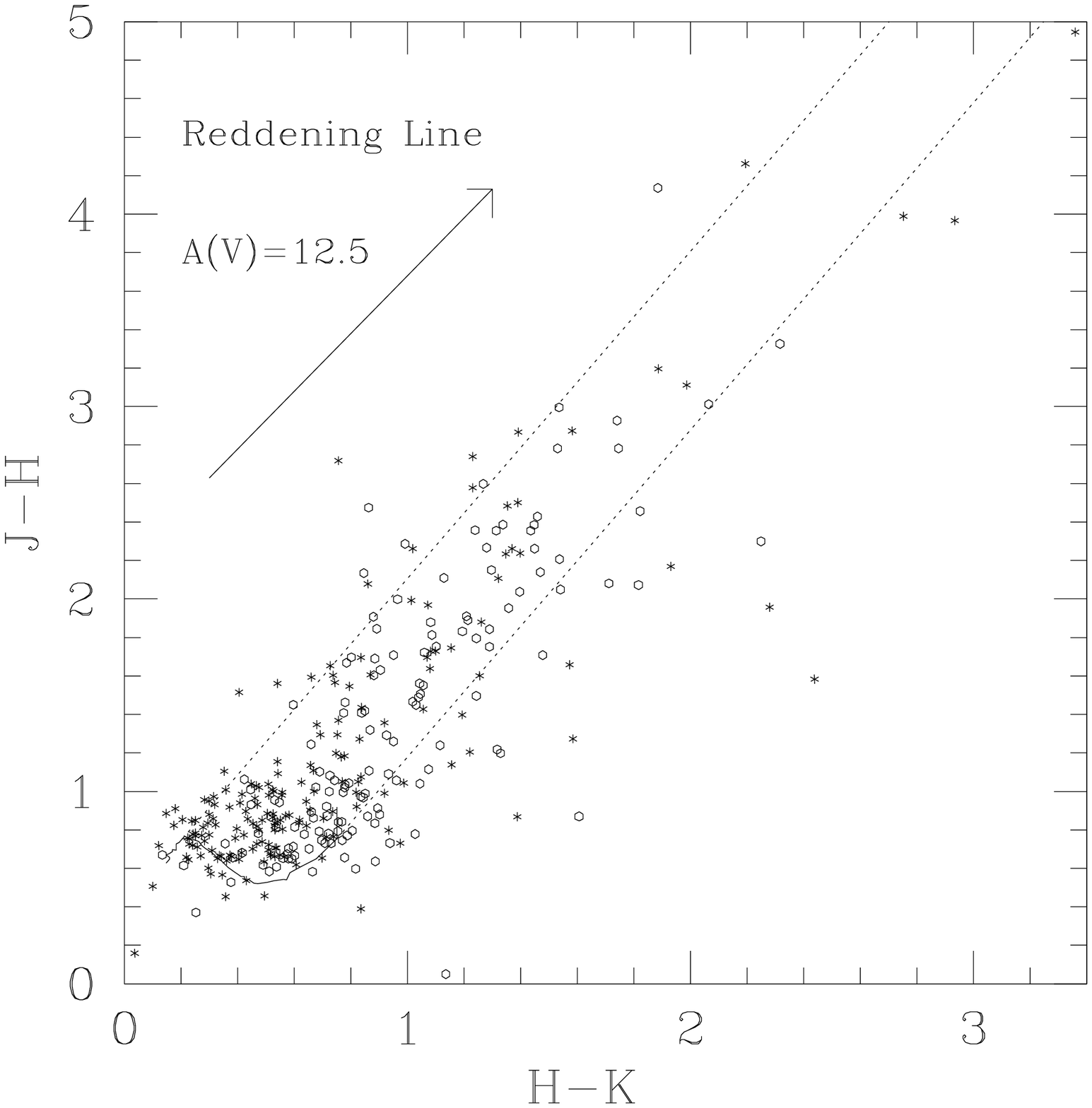}}

\end{picture}
\end{center}

Figure 9. Two Colour Diagram. More luminous sources with dereddened H mag
H$_{dr}<13$ are marked with stars. Less luminous sources (H$_{dr}>13$) are
marked with open circles. The combined isochrone described in section 3.1 
is plotted as a solid line. Dashed lines parallel to the reddening vector
are drawn from the extreme tangents to the isochrone and enclose most of the 
data points.
\end{figure*}

Since we do not have effective temperatures for most PMCs, an age must be
assumed for the sources in order to derive a mass from the
luminosity using a theoretical isochrone. In Figure 8 we investigate possible 
forms of the IMF for different samples and assumed ages. Since the colours and 
fluxes of the theoretical isochrone become unreliable below 1800~K (see section 3.1) 
this conversion becomes very uncertain for masses below 5~M$_{Jup}$ and we somewhat
arbitrarily use the dereddened H band fluxes for the luminosity to mass conversion.
However, given the binning adopted in Figure 8 this has little consequence for our
analysis.

The bottom panels of Figure 8 show the mass function for subsamples with A$_V<5$, 
in order to reduce background contamination and ensure good completeness. The top panels
include the full range of measured extinctions (calculated from the colour magnitude 
diagrams as described at the end of Section 3.1) and are incomplete at low masses
due to brown dwarfs being reddened beyond the detection threshold.

In the A$_V<5$ plots a dashed
extension to some of the least massive bins corrects for the effect of incompleteness
due to limited sensitivity, under the assumption that the typical extinction is
A$_V=2.5$. In the left panels we 
assume an age of 1~Myr and in the right panels we assume an age of 10 Myr. 
Most previous age estimates for cluster members, made using the HR diagram or 
detection of circumstellar matter, have found a typical age of slightly less
than 1~Myr. In a recent spectroscopic survey Slesnick, Hillenbrand \& Carpenter (2004)
suggested that the M-type cluster population has a much broader age distribution than
the earlier spectral types, including a substantial fraction ($\sim 50\%$) of 10~Myr old 
sources whose masses are much higher than would be derived from a simple luminosity to mass 
conversion using a 1~Myr 
isochrone. The status of these apparently warmer, older and more massive M type sources 
is presently unclear: in observations of the outer parts of the cluster as covered in this 
survey the fraction of luminosity selected brown dwarfs with spectra earlier than mid-M type 
found in the optical spectra of 
Riddick, Roche \& Lucas (2005) and the H band spectra of Lucas et al.(2001) is closer to 10\%.
In addition the L band photometry of Lada et al.(2004) in the central regions of the
Trapezium yielded a similar disc fraction in stars and brown dwarfs ($\sim 50\%$) in both 
photometrically selected brown dwarfs and a spectroscopically confirmed subsample. This
result is hard to reconcile with a large 10~Myr old population, given that disc lifetimes 
appear to be similar in stars and brown dwarfs, eg. Mohanty et al.(2005). 

However, it is known that there is a diffuse population in Orion with ages of 
$\sim 10$~Myr on spatial scales up to a degree
from the Trapezium cluster (eg. Hillenbrand 1997) and while this population
is expected to introduce $<1$ spurious PMC in the survey region (see Lucas et al. 2001) 
it is useful to consider the effect that a 10~Myr age would have on the results.

The 1~Myr histograms shows a mass function which peaks near 0.2~M$_{\odot}$ and
declines through the brown dwarf regime, in agreement with previously published surveys.
The population of PMCs is small, confirming the impression given by the colour
magnitude diagrams and the K band luminosity function. The mass function in the lower
left panel is not well fitted by a power law but since power law indices have the virtue
of simplicity we provide one for that panel here. Using the Chi squared minimisation straight 
line fitting routine LINFIT in the IDL package and adopting Poisson errors we find, in the 
$log(M)=-1.0$ to -2.2 range, that the index is $\alpha = 0.31 \pm 0.11$, where 
$N(M) \propto M^{\alpha}$. 

It should be noted that there is strong mass segregation in the Trapezium cluster. 
Hillenbrand (1997) found that the ratio of high mass to low mass stars is fairly constant at
distances $r \ge 0.3$~pc from the cluster centre but there is a much larger fraction
of high mass stars nearer the cluster centre. More than 90\% of our survey area lies
at $r \ge 0.3$~pc from $\theta_1$~Ori C, so we can assume that the derived IMF for
A$_V<5$ sources provides information for a representative sample of the outer cluster 
provided that (i) there is no significant change in the IMF with increasing extinction
along the line of sight, and (ii) the radial IMF variation reported by Hillenbrand (1997)
holds true for substellar sources also.

The $\alpha = 0.31$ value indicates a slow decline in the IMF at substellar masses 
which agrees well with the results reported for the $\rho$~Ophiuchus and IC348 clusters 
(Luhman et al. 2000; Muench et al. 2003). It contrasts with the value $\alpha = -0.8 \pm 0.4$ 
reported in the $\sigma$-Orionis cluster (Bejar et al. 2001) and the large number of substellar 
candidates detected by Oasa (2003) in the Chamaeleon I, S106 and NGC1333 regions.
However these quoted surveys are all based on photometric candidates and in the more
diffuse clusters the amount of contamination by background stars is hard to estimate 
without spectroscopic follow up of a large sample. Hence, it is not clear at present 
whether there are large variations in the substellar IMF between star forming regions, 

If PMCs are excluded from 
consideration then brown dwarf candidates in the $0.012<$M$<0.075$M$_{\odot}$
mass range constitute 28\% of the A$_V<5$ subsample and 32\% of the sample that includes all 
extinction values. This difference, which is visible in the left panels of Figure 8,
suggests that there is some background
contamination among the more highly reddened candidates, as noted in section 3.2. 
This is consistent with the detection of 2 possible background stars with A$_V \approx 10$
in the sample of 21 low mass sources in the spectroscopic sample of Lucas et al.(2001). 
However, the effect may be slightly exaggerated by imperfections in the colours of the 
isochrone in Figure 2: there is a small gap of $(J-H) \approx 0.1$ (or A$_V \approx 1$) 
between the isochrone and the data points in the brown dwarf mass range that does not exist 
at stellar or planetary masses (see also Figure 6). Hence the extinction may be slightly 
overestimated for the 
brown dwarfs, reducing the number in the A$_V<5$ subsample by up to $20\%$. This is 
sufficient to eliminate the 4\% discrepancy in the brown dwarf fraction without invoking
background stars. The extinction histograms in Figure 6 and the analyses of 
Muench et al.(2002) and Hillenbrand \& Carpenter (2000) suggest that a small amount of 
contamination is expected among the fainter brown dwarf candidates with K$ \ge 16$, so a 
compromise brown dwarf fraction of 30\% is perhaps the best estimate in this survey
of the outer cluster.

With an assumed age of 1~Myr the A$_V<5$ subsample (lower left panel of Figure 8)
is almost 100\% complete to M$=5~$M$_{Jup}$ and 75\% complete in the bin at $log(M)=-2.4$ 
(3-5~M$_{Jup}$). Both the 
A$_V<5$ subsample and the full sample show an apparent rise in the bin at $log(M)=-2.4$, 
which is very likely due to background contamination. The analysis of Muench et al.(2002) 
illustrates how a background population with a Gaussian distribution of A$_V$ (for example)
can introduce 
significant contamination among the PMCs with A$_V<5$ even if the median A$_V \ge 10$, 
owing to the large number of background sources expected at K$>17$. At planetary masses the 
difficulty of detecting highly reddened sources ensures that most of the detected sources 
are in the A$_V<5$ subsample. In this subsample of 208 sources there are 24 PMCs. The correction
for incompleteness in the $log(M)=-2.4$ bin (assuming typical A$_V=2.5$ in the subsample) 
raises this to 29 PMCs. Neglecting the single PMC in the $log(M)=-2.6$ bin these 
constitute 13.2\% 
of the total. This may be regarded as an upper limit to the PMO (planetary mass object) number 
fraction in the mass range $0.0032<$M$<0.0120$M$_{\odot}$, or 3-13~M$_{Jup}$. 
However, the A$_V<5$ mass function appears to drop by a factor of two at the deuterium burning 
limit: there are 14 sources per bin in the $log(M)=-1.6$ and -1.8 bins but only 6-7 in 
the $log(M)=-2.0$ and -2.2 bins. This measured drop deviates by 2.8-$\sigma$ from a flat mass 
function with a mean of 14 sources per bin. If the mass function for bona fide PMOs is flat at 
6.5 sources per bin then the PMO number fraction in the $0.0032<$M$<0.0120$M$_{\odot}$ mass range 
(constituting slightly less than 3 bins) is 9.5\%. It should be emphasised that these masses
are only notional, being subject to significant but poorly quantified theoretical uncertainties 
(see Baraffe et al. 2002) for sources with ages of $\sim 1$~Myr. There is some recent evidence
that pre-main sequence isochrones are reliable both at stellar masses (Hillenbrand \&
White 2004, based on dynamical mass measurements of binaries) and at substellar masses
(Mohanty et al. 2004a, 2004b, using masses estimated from surface gravity measurements for
brown dwarfs with ages of a few Myr.)

The 10~Myr histogram is shifted to higher masses by a factor of 2.5 to 3 relative to the 
1~Myr plot. However, even under this extreme assumption the A$_V<5$ population includes 
a few PMCs and more than 20 very low mass brown dwarf candidates with M$<0.02$M$_{\odot}$.

\subsection{K band Excesses?}

Muench et al.(2001) reported the detection of K-band excesses attributable
to hot circumstellar dust in $\sim 65\%$ of Trapezium brown dwarfs.
Lada et al.(2004) confirmed and extended this result by including L band 
photometry and a spectroscopically selected sub-sample for which the 
photospheric colours could be more reliably established. This result was
somewhat unexpected since theoretical modelling (eg. Liu et al. 2003) of a 
conventional circumstellar disc in hydrostatic 
equilibrium predicted that infrared excesses would be smaller
in brown dwarfs than in stars. Further interest in the question of the
matter distribution around brown dwarfs stems from: (a) the hypothesis of brown 
dwarfs as ``ejected stellar embryos'' (Reipurth \& Clarke 2001; Bate, Bonnell
\& Bromm 2003) which would be expected to have at most small discs; and (b)
the calculation by Walker et al.(2004) that young brown dwarfs with large discs 
(r$\ge 150$~AU) could be hidden by the large flaring angle and hence remain 
undetected.

\begin{figure*}
\begin{center}
\begin{picture}(200,370)

\put(0,0){\includegraphics{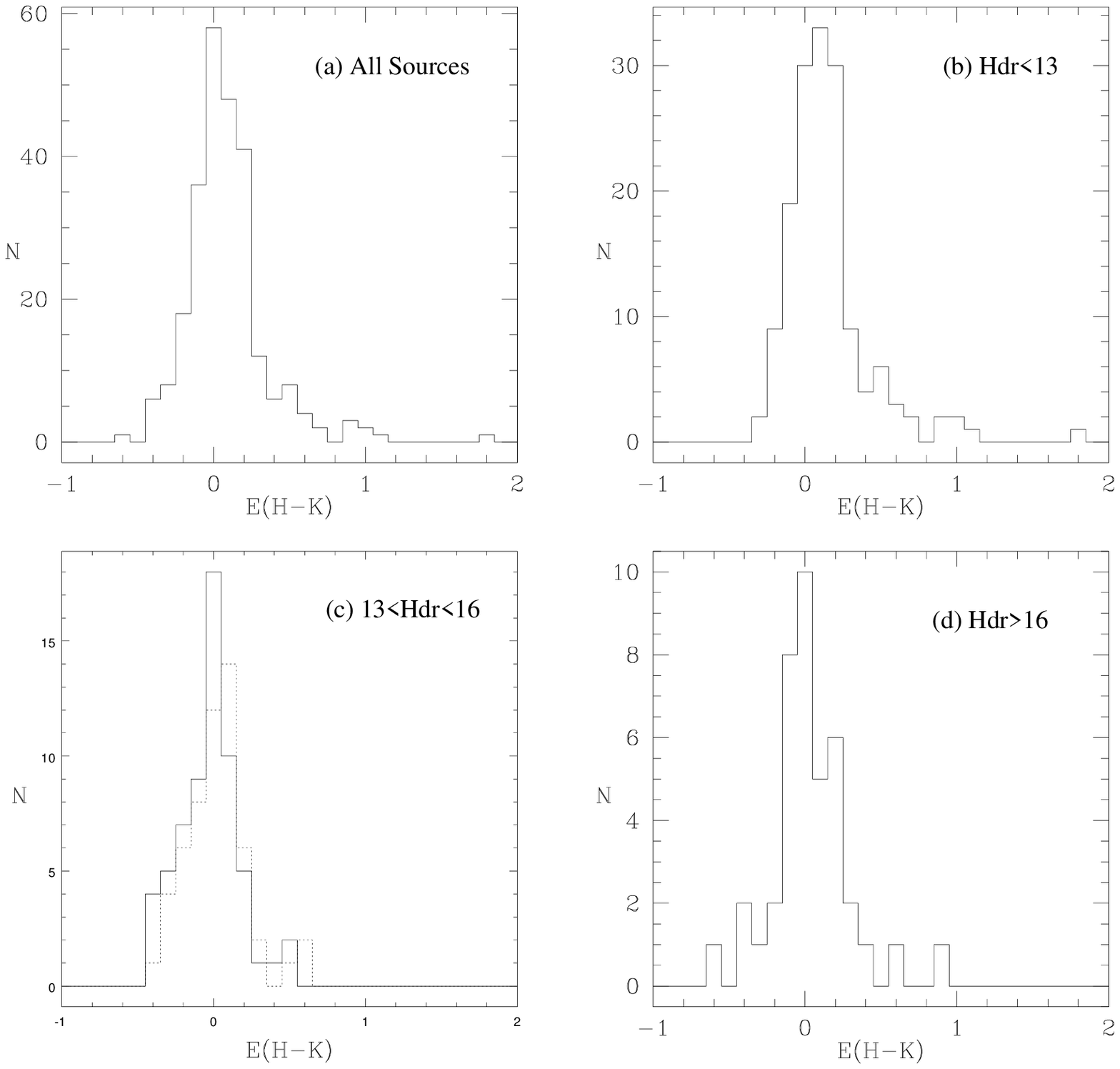}}

\end{picture}
\end{center}
Figure 10(a-d). Histograms of H-K colour excesses, after dereddening
to the 1~Myr isochrone. (a) all sources; (b) H$_{dr}<13$; 
(c) $13<$H$_{dr}<16$; (d) H$_{dr}>16$, where H$_{dr}$ is the dereddened H magnitude.
\end{figure*}

In Figure 9 we plot all JHK detections in the survey in a two colour diagram.
The intrinsic colours for 1~Myr old sources are illustrated by the curved line,
using the combined isochrone described in Section 3.1, incorporating field dwarf 
colours for the reddest objects at T$_{eff}<1900$~K.
The two reddening vectors extending from the extremes of the isochrone 
enclose the majority of the data points, showing that this type of diagram alone
is of little value in identifying sources with (H-K) colour excesses owing to 
the wide range of possible (H-K) colours in M and L-type dwarfs. Previous workers
have tended to truncate the isochrone at M9, corresponding to (H-K)=0.5, but
in this very deep survey L-type sources may well be included. Among the sources
with low extinction ((J-H)$<1.2$) the fainter sources with dereddened H mag 
H$_{dr}>13$ (open circles) tend to have have redder (H-K) colours than the brighter 
sources (stars), most likely because fainter sources have cooler and redder 
photospheres. The two colour 
diagrams for the individual fields are similar (not shown) except that Field 3 
has more scatter, providing a large fraction of both the sources with apparent (H-K) 
excess and anomalously blue colours (nearly all of stellar mass) lying above and to 
the left of the reddening lines. Although all 3 fields lie at similar distance from 
the cluster centre, Field 3 is closest to the densest parts of the OMC-1 ridge, 
associated with ongoing star formation. The wider scatter in the colours might therefore 
be explained by the presence of younger sources with scattered light (blue sources) 
and emission from hot circumstellar dust (red sources). 

Lada et al.(2004) also observed (in the central regions of the 
Trapezium) that some luminosity selected brown dwarfs run contrary to the general 
trend and have colours bluer than M6 (H-K=0.5), which is confirmed here. They 
suggested that these low luminosity blue sources might be related to the unexpectedly 
warm low luminosity sources reported by Slesnick, Hillenbrand \& Carpenter (2004), 
whose nature is presently unclear. These works show that there is considerable 
difficulty in measuring K band excesses caused by hot circumstellar dust in a two colour 
diagram, due to both uncertainty in intrinsic colours of the photosphere and the 
effects of scattered light. It is clearly preferable to observe at longer wavelengths, 
where infrared excesses are larger and scattered light is less important. Samples with 
measured spectral types also improve confidence in the analysis (eg. Liu et al. 2003; 
Lada et al. 2004). 

In Figure 10(a) we show the histogram of (H-K) colour excesses for 256 point sources in 
our sample with well measured J, H and K band fluxes and M$<1.2$M$_{\odot}$ 
(excluding the T dwarf candidate described in the next section, which has a very 
large negative excess). (H-K) excesses are calculated by measuring the excess after 
dereddening to the H vs.(J-H) 1~Myr isochrone (with field dwarf colours at 
T$_{eff}<1900$~K). Specifically, E(H-K)=(H-K)$_{obs}$ - (H-K)$_{ext}$ - 
(H-K)$_{int}$, where (H-K)$_{int}$ is the intrinsic colour of the photosphere
and (H-K)$_{ext}$ is the change in the colour due to extinction.
The distribution has a peak at 0.0 magnitudes and a fairly symmetric bell
shaped appearance, consistent with an approximately Gaussian distribution of
photometric errors. There is a slightly larger number of sources with a large 
positive excess (E(H-K)$>0.5$~mag) than a large negative excess but there is little
sign of the large fraction of sources with modest excesses (E(H-K)$ \sim 0.2$) 
that would have been expected based on the previous investigations quoted above. 

Figures 10(b-d) break down the diagram as a function of dereddened luminosity.
Figure 10(b) indicates that sources of stellar mass do tend to have K band excesses
(median 0.1 mag). This histogram is composed of approximately equal amounts
of data from this survey and data from other sources (owing to saturation
of bright sources in the Flamingos data) but the distributions of (H-K) excess for 
the two components appear to be identical (not shown).

Figures 10(c-d) show that there is little sign of (H-K) excess among the less luminous 
sources. The solid histogram in Figure 10(c) gives the result from sources with 
dereddened H mag $13<H_{dr}<16$ using the intrinsic colours of the theoretical 
isochrone. This histogram suggests a tendency for the excesses of luminosity
selected brown dwarf candidates to be slightly negative. Since this is not seen
at stellar masses it cannot be explained by errors in the photometric zero points.
However it may be explained by a possible small error in the (J-H) colours of the 
isochrone, as noted in sections 3.1 and 3.3. The dashed histogram
overplotted in Figure 10(c) shows that the negative excesses are eliminated if 
we adopt slightly redder (J-H) colours for the isochrone, thereby reducing the 
calculated extinction toward each source and increasing E(H-K). The dashed histogram
was constructed using the main sequence dwarf (J-H) colours adopted by Muench et 
al.(2003, figure 2) for the isochrone of D'Antona \& Mazzitelli (1997), which are 
0.05 to 0.13 mag redder than the isochrone on the MKO system in the appropriate 
magnitude range. The (H-K)$_{int}$ colours of the NextGen isochrone were retained 
for the calculation.

The dashed histogram in Figure 10(c) shows only a small tendency for E(H-K) 
to be greater than zero. This is surprising given that the works of Lada et 
al.(2004) and Muench et al.(2001) found a large fraction of (H-K) excesses among 
both stars and brown dwarfs in the central parts of the Trapezium cluster, despite the 
fact that Muench et al.(2001) used the colours of the NextGen isochrone. 
Comparison of Figure 1 of Muench et al.(2001) with our corresponding colour magnitude 
diagrams (Figures 2 and 3) shows that the distribution of (H-K) colours is similar 
in the two surveys. However, the Muench et al. dataset contains a large number of
brown dwarf candidates which have bluer (J-H) colours than the NextGen isochrone, in 
contrast to our data where brown dwarfs tend to appear redder than the isochrone.
Errors in the photometric zero points are not a likely explanation since the colours
of the datasets agree well for brighter sources and the Gemini zero points have been
checked against those of Muench et al.(2002), see section 2.
A possible explanation is that Muench et al. found many of their sources
with blue (J-H) colours to be ``proplyds'' (O'Dell \& Wen 1994) and these tended
to have large (H-K) excesses. LR00 noted that proplyd sources appear to have slightly 
bluer (J-H) colours than others, probably due to the effects of scattered light. 
This effect may be more obvious among low mass sources due to the relatively large scale
height of brown dwarf disks (Walker et al. 2004). Proplyds would be expected to have
relatively large (H-K) excesses both because the distortion of their (J-H) colours
leads to an underestimate of (H-K)$_{ext}$
and because of the confirmed presence of large amounts of circumstellar matter.
The rarity of (H-K) excesses and sources with blue (J-H) colours in the Gemini data 
may indicate that there is both less hot dust in the outer parts of the cluster
and fewer proplyd structures (circumstellar envelopes). This would imply a slightly
greater typical age for sources in this survey.

Owing to the strong sensitivity of the E(H-K) histograms to small changes in colour 
the above explanation may well be an over interpretation of the data.
The width of the distributions in Figure 10 is likely to be influenced 
by source variability in addition to photometric scatter, so the dataset
is not ideal for detection of (H-K) excesses. 

For the faintest sources (Figure 10(d)) the histogram is fairly symmetric about
zero. Figure 10(d) is likely to be influenced by the contribution of
background stars (see section 3.2) but we note that background stars would need
to be quite cool ((H-K)$_{int} > 0.52$) in order to avoid displaying spurious (H-K) 
excesses.

Finally, we consider the effect of different extinction laws on the analysis.
The reddening law adopted in this paper is that of Rieke \& Lebofsky (1985), for
which the gradient of the reddening line $S=(A_J-A_H)/(A_H-A_K)=1.70$. The analysis
of Muench et al.(2001) used the extinction law of Cohen et al.(1981), which yields
a similar gradient of $S=1.69$, despite a somewhat lower ratio of infrared to optical 
extinction. By contrast the $\lambda^{-1.61}$ extinction law of Cardelli, Clayton \& 
Mathis (1989), which we adopted in LR00, has $S=1.50$. A slope of $S=1.50$ appears 
to be inaccurate, since it leads to a tendency for sources with high extinction to 
have smaller or negative (H-K) excesses. 

\begin{figure*}
\begin{center}
\begin{picture}(200,250)

\put(0,0){\includegraphics{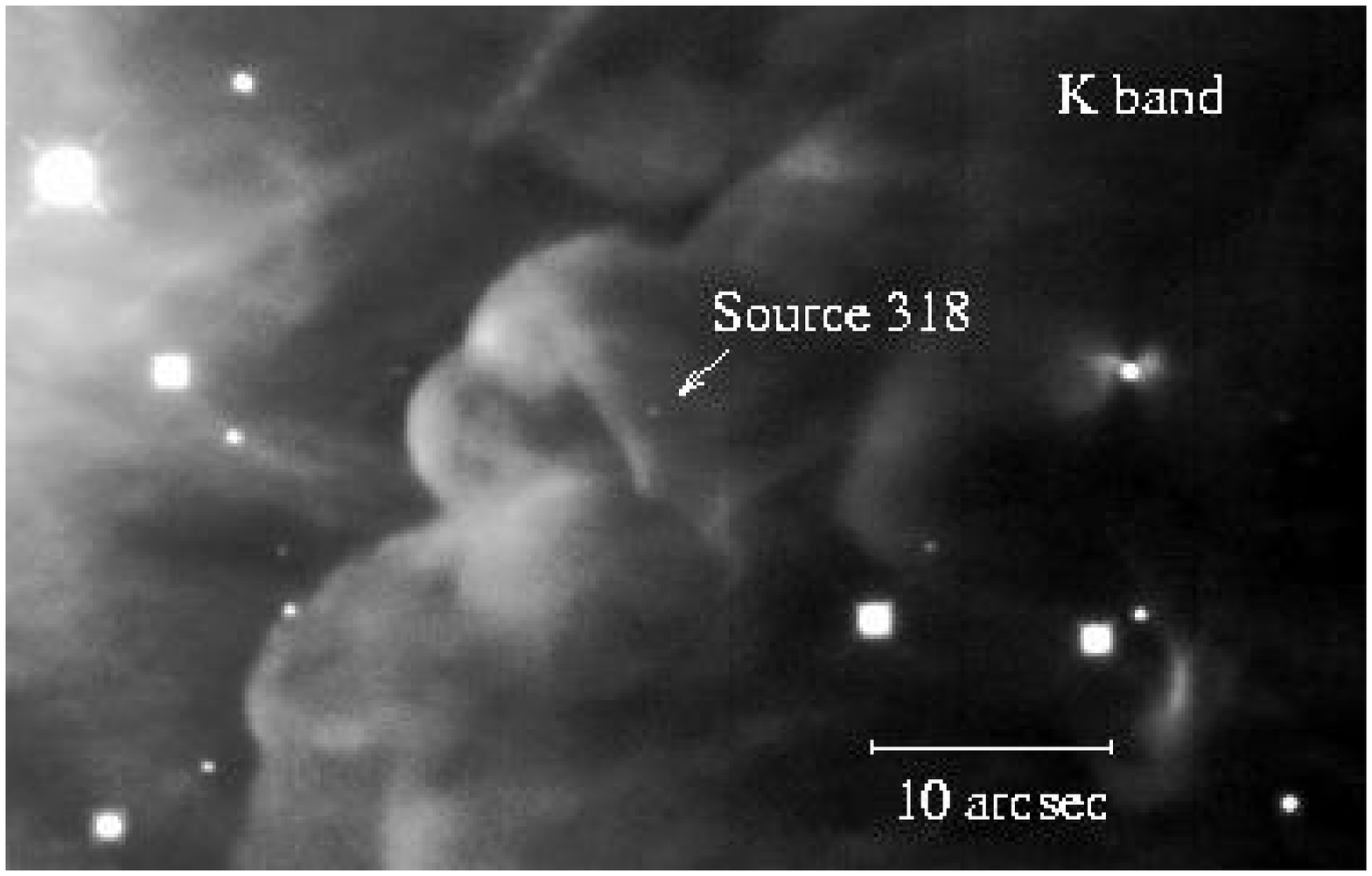}}

\put(0,0){\includegraphics{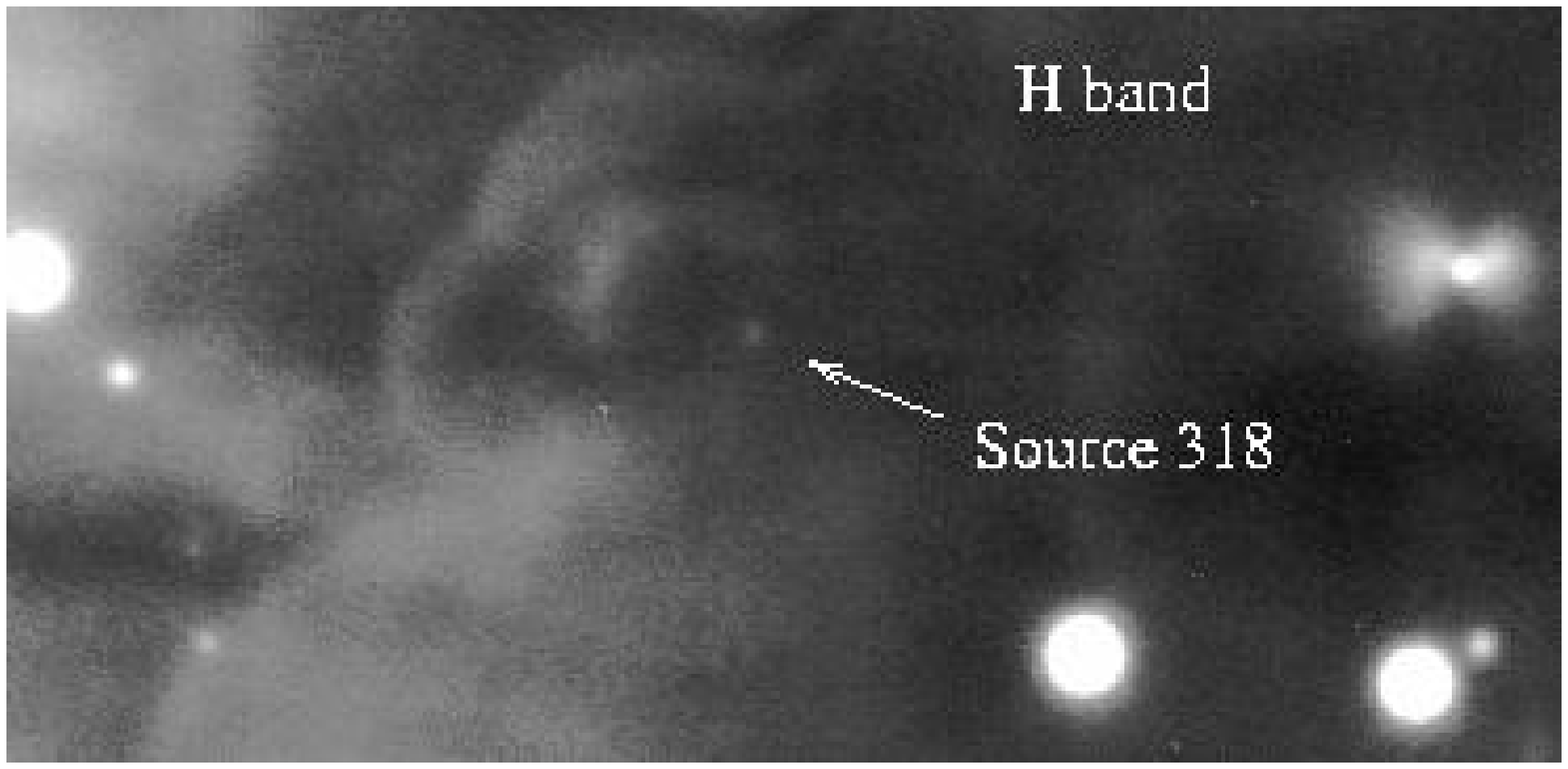}}

\put(0,0){\includegraphics{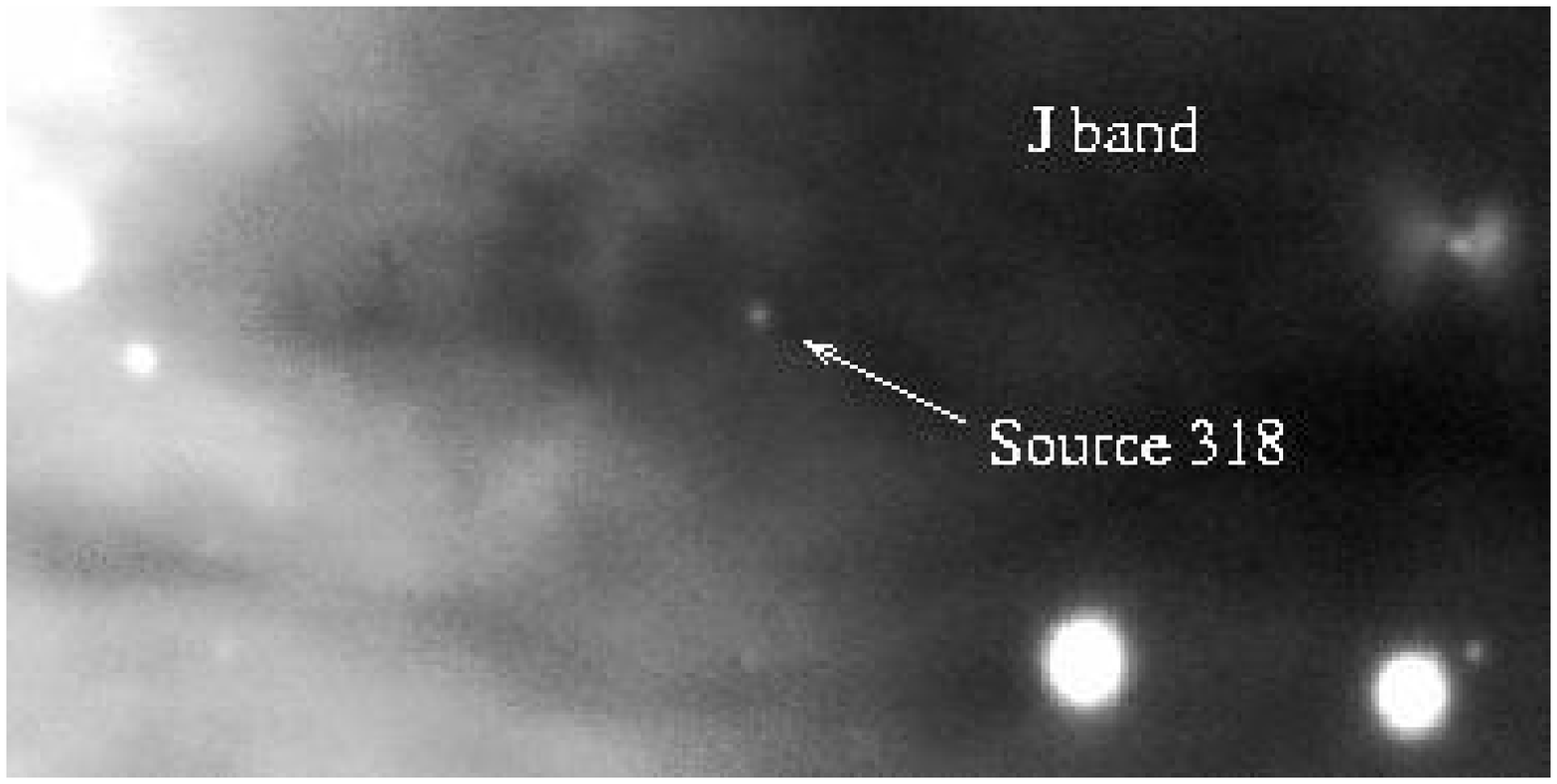}}

\end{picture}
\end{center}
Figure 11. Images of Source 318, a T dwarf candidate (left) at K band; (right) close up
views in the H and J bands. Many of the nebulosity features seen adjacent to
Source 318 at K band are unseen at J and H, suggesting that the source lies in front of a 
region of high extinction. North is up and east is to the left in these images and all
later images in this paper.
\end{figure*}

\subsection{A T Dwarf Candidate}

In section 3.1 we noted that a single faint source, Source 318, has very blue 
colours, lying well to the left of the theoretical isochrones in Figure 2 and 3.
The source colours are (J-H)=0.37, (H-K)=0.25, and the fluxes are K=18.95, H=19.21,
J=19.58. The location of Source 318 (or 071-307 using the coordinate based naming 
system) is indicated in Figure 11. It lies in a region where many nebulous 
structures are present 
(some with the appearance of bow shocks) and the source density is high, 
suggesting that there has been recent star formation. 
If the object is a cluster member then the luminosity derived mass
from the 1~Myr DUSTY isochrone would be M=2.5 to 4.5$~M_{Jup}$, depending 
which filter is used for comparison. The corresponding temperature would be 
1500 to 1800~K. Taken at face value, the blue colours imply a lower temperature, 
corresponding to an early T-type dwarf with significant methane absorption and 
T$\sim 1400$~K. Such a low temperature could be 
explained if Source 318 is younger than 1~Myr and has a mass of perhaps 
2 to 3$~M_{Jup}$. (The DUSTY isochrone does not extend below 1~Myr, for the 
very good reason that prediction becomes increasingly unreliable at these very young ages,
see Baraffe et al. 2002). The low temperature would then be counterbalanced by a
larger size, to produce the observed fluxes. 

However several other explanations for the data must be considered. These are:
\begin{itemize}
\item variability. The apparently blue colours could be an illusion caused 
by variations in source flux between the observations in different filters.
As noted in section 2, variability at the level of 0.1 to 0.2 mag is common in low 
mass Young Stellar Objects (YSOs), so it is certainly possible that the true colours are 
redder than our photometry indicates. However, Source 318 is the only faint source 
in Figure 2 with well measured fluxes that is so blue in (J-H), suggesting that this 
type of effect on the colours is rare. The blue (H-K) colour then requires
an additional variation in flux, which makes it a low probability event
despite the lare number of sources in the survey. If variability is the explanation 
for the source colours this would tend to support cluster membership.

\item photometric scatter. The internal photometric uncertainties for this source
are small ($\sim 0.1$ mag) but owing to the presence of nebulosity gradients,
particularly in the K band, it is conceivable that the blue colours are a 
consequence of rare 2 or 3-$\sigma$ photometric errors. This explanation is 
unlikely for the same reason as given above: it would require two low probability
events. If this has occurred then Source 318 could be a cluster member, a background star
or even a distant unresolved galaxy. We noted in Section 2 that there are significant 
uncertainties in the photometric zero points at a level of up to 0.1 mag. However these 
would systematically affect all sources in each field, so this cannot help to explain the 
colours. 

\item a foreground T-dwarf. The absence of reddening, despite the presence of 
strong extinction towards adjacent nebulosity features shown in Figure 11,
indicates that Source 318 is very unlikely to be a background star. However, its
fluxes and colours are consistent with an early T dwarf (most consistent with
type T3) located in the foreground
at a distance of $\sim 130$~pc (using data from Knapp et al. 2004). The T-dwarf space 
density in the 800 to 1400~K range is believed be 0.01 to 0.06~pc$^{-3}$
(Burgasser 2004) and the volume of space at d$<130$pc in our 26~arcmin$^2$ survey 
area is 1.5~pc$^3$. Including the effect of completeness and neglecting the
(unknown) scale height of the T dwarf distribution in the plane, the probability of 
finding any type of T dwarf in our 26~arcmin$^2$ survey area down to a limit of 
K=19.7 (a factor of 2 fainter than Source 318 and approximately our 50\% 
completeness limit) is in the range 0.02 to 0.11. Hence, while the possibility of a 
foreground T dwarf cannot be ignored it is another low probability event.

\item scattered light. In LR00 we discovered that $\sim 14\%$ of Orion sources
had anomalously blue (I-J) colours and correlated closely with the presence of 
circumstellar matter (the 'proplyds' of O'Dell \& Wen 1994). It is likely
that scattering by circumstellar matter is at least partly responsible for these
blue colours, although the corresponding effect on (J-H) colours appeared to be
smaller. A few of the sources in Table 2 have circumstellar matter that is marginally
resolved in 1 of the 3 filters and these sources tend to have anomalous colours. 
Hence, scattering by unresolved circumstellar matter is perhaps the most likely
alternative explanation for the blue colours of Source 318.
\end{itemize}

It is clear that spectroscopic follow up will be required to establish the
nature of the source. Detection of methane absorption should be well within
the capability of planned observations with Gemini South. This would refute
most of the alternative explanations but detection of low surface gravity
features would be required to rule out a foreground T dwarf and that will be more
difficult. 

In summary, Source 318 certainly merits further scrutiny since its
possible very low mass (perhaps 2-5~M$_{Jup}$) challenges the theory of opacity 
limited star formation and unlike the other very low luminosity candidates in the 
survey we consider it very unlikely to be a background star.

\subsection{Spatial Analysis and Binarity}

\begin{figure*}
\begin{center}
\begin{picture}(200,200)

\put(0,0){\includegraphics{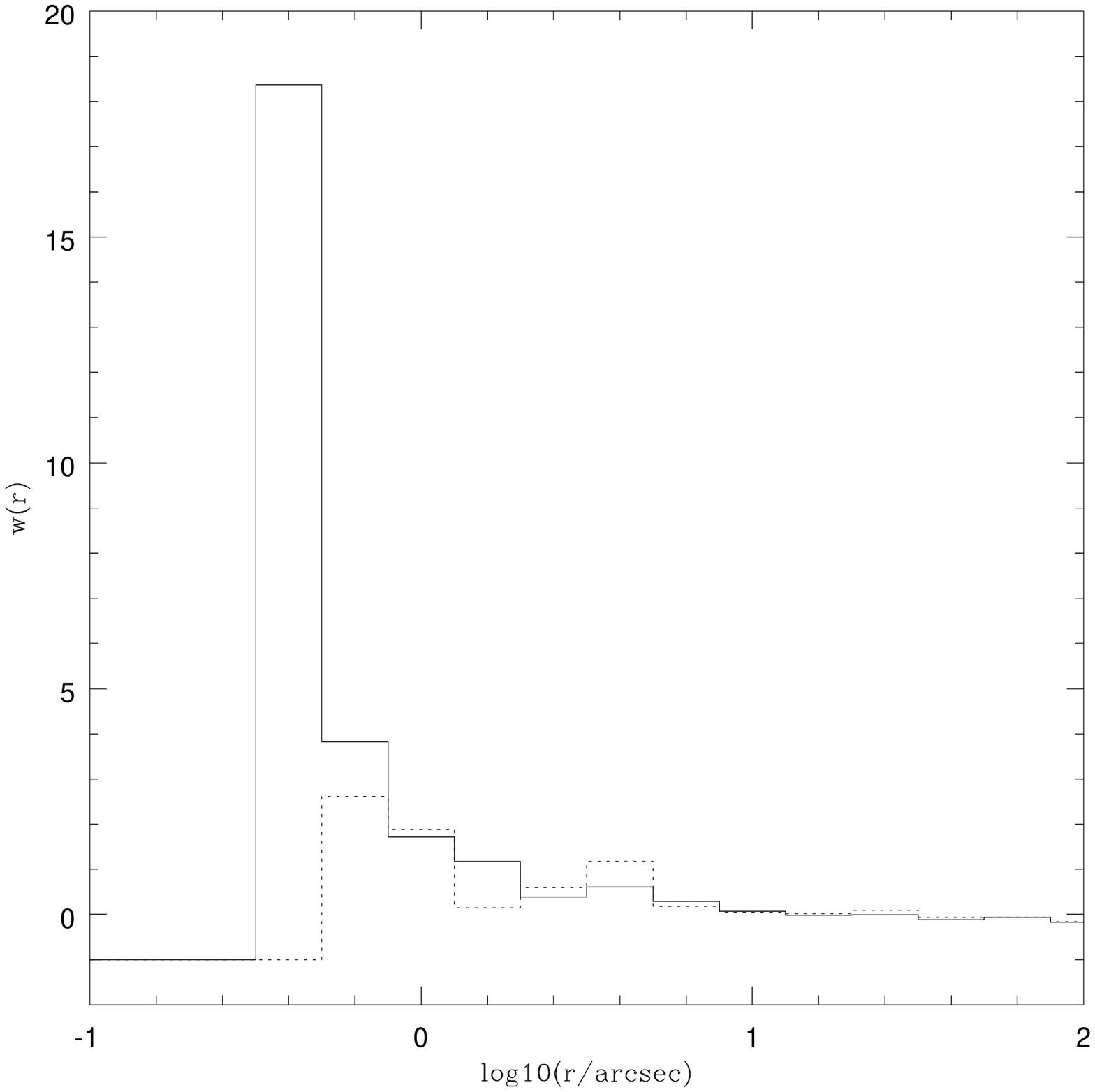}}

\end{picture}
\end{center}
Figure 12. The Two point correlation function for the Gemini sources. (a) stars 
(solid histogram); (b) brown dwarfs (dashed histogram). A random distribution of sources
would have w($\theta$)=0 at all separations. A spike due to binarity is seen at small 
separations in the stellar subsample but only a small rise at small separations is seen
in the brown dwarf subsample. This plot does not correct for the selection effect caused
by reduced sensitivity to close binaries.
\end{figure*}

We conducted an analysis of binarity and clustering in the stellar and brown dwarf 
populations to search for differences which might shed light on the brown dwarf formation
mechanism. Previous smaller scale studies of binarity in more central regions of the
Trapezium cluster (eg. Simon 1997; Petr et al. 1998; Prosser et al. 1994) have 
indicated that the incidence of binarity in the stellar mass population is much lower 
than in nearby low mass star forming regions such as Taurus-Auriga and $\rho$ 
Ophiuchus but is comparable to the field population at the measured separations 
of a few dozen to a few hundred AU. Simon (1997) reanalysed the HST/WFPC1 dataset of
Prosser et al.(1994) (containing 319 stars) and showed that the mean surface density of 
companions (MSDC) as a function of separation is identical to that of the Taurus-Auriga 
and $\rho$ Ophiuchus star forming regions at separations up to $\sim 300$~AU but the 
binary distribution is terminated at that point. At larger separations the slope of the 
MSDC function becomes much flatter in the Trapezium cluster, corresponding to the 
changeover from binarity into the regime of random distribution or clustering, a 
phenomenon first discovered by Larson (1995) in Taurus-Auriga. Hence the higher incidence
of binarity in Taurus-Auriga and $\rho$ Ophiuchus is apparently due only to the presence
of wide binaries (up to several thousand AU) in those regions.

It is very likely that the termination of the binary distribution at separations of a few
hundred AU is due to frequent close encounters with other stars on that scale within 
a few $\times 10^5$~yr in the central regions of the Trapezium Cluster 
(eg. Petr et al. 1998; Bate, Clarke \& McCaughrean 1998; Kroupa, Petr \& McCaughrean 1999). 
After analysing a sample of 45 stars Petr et al.(1998) suggest that the incidence of 
binarity may be lower among Trapezium stars of mass M$<1.5~$M$_{\odot}$ than among more 
massive stars. They found no sign of brown dwarf companions to more massive stars 
down to a sensitivity limit of K$\approx 16$, although 4 very close stellar binaries were 
detected.

\subsubsection{Gemini data}
The two point correlation functions for stars and brown dwarfs are
plotted in Figure 12, correcting for boundary effects with method 3 of those 
described by Bate et al.(1998). For each subsample (stars, brown dwarfs) every
source in the subsample is correlated with every other source detected in the survey 
region. Hence stellar binaries contribute twice to the stellar function, star/brown 
dwarf binaries contribute once to each function, and brown dwarf binaries contribute
twice to the brown dwarf function. 
The spike in the correlation function for stars in the 2 bins at the smallest measured 
separations $\theta=$0.32 to 0.79 arcsec ($140<r<350$~AU)
shows that there is a statistically significant population of binaries 
which cannot be explained by chance alignment. (The two point correlation
function $w(\theta)$=0 for a random distribution of sources within the survey
area and is related to the MSDC by the equation 
$w(\theta)= (MSDC(\theta)/(N_{*}/A)) - 1$, where $N_{*}$ is the number of stars in 
the survey and $A$ is the survey area (Peebles 1980). It is notable that binaries appear 
less common in the brown dwarf population (see dashed histogram in Figure 12), 
particularly in the smallest measured bin at $log(\theta)=-0.4$. Sensitivity to binaries is 
similar in both histograms, being essentially independent of flux. At separations 
$\gtsimeq 1$ arcsec there is a significant probability of chance alignment causing 
spurious binaries (see eg. Petr et al. 1998), so we limit our discussion of binarity 
statistics to the 2 closer bins.
Selection effects have been minimised by excluding PMCs from the brown dwarf sample, 
since these are close enough to the sensitivity limit for companions to scatter below 
it and they are also likely to be significantly contaminated by background stars. 
For the same reasons the subsamples exclude brown dwarf candidates and stars with 
H mag $>17.19$, which is the magnitude of a 1~Myr old source at the deuterium burning 
threshold with zero extinction. 
In practice none of these excluded faint sources are found to be in binary systems. 
The subsamples also exclude sources at the edges of the data mosaic, for which sensitivity 
to faint brown dwarf binaries is poor. It 
should be noted that there are strong observational selection effects against the detection 
of the closest binaries which reduce the height of the spike in the plotted two point 
correlation function. Close binaries which are detected tend to be near equal 
luminosity pairs located in the part of the survey with the highest spatial resolution 
(Field 3). We do not attempt to correct for this selection effect since the 
companion luminosity function is unknown.

This difference between stellar and brown dwarf two point correlation functions has a
rather modest statistical significance. The spike in the stellar correlation function 
in the $log(\theta/arcsec)=$-0.4 and -0.2 bins is due to 6 stellar binary systems (see
Table 3) and 1 star/brown dwarf pair from a subsample of 205 stars, a fraction reasonably 
consistent with the results for the cluster centre referenced above, given the small
range of separations considered here.  
The brown dwarf subsample in Figure 12 numbers 56 sources but includes only half of 1 binary 
system: Source 216 is located in an apparent star/brown dwarf pair ($\theta=0^{''}.72$) and 
is the only high probability member of a binary system in the 2 closest populated bins. 
The expected number of apparent binary pairs in a sample
of size N caused by chance alignment is $N \pi \theta^2 \Sigma$ (see Petr et al.1998), 
where $\Sigma=3.94 \times 10^{-3}$arcsec$^{-2}$ is the surface density of all sources 
in the survey. 
$\Sigma$ is fairly uniform in the survey area as indicated by the fact that 
$w(\theta)\approx 0$ at $\theta >10$~arcsec. We would expect to observe a decline in 
source density with increasing distance from the cluster centre (see Bate et al. 1998)
but the effect is not very significant in this survey region.
For an ideal survey sensitive to all possible pairings the expected contamination is 
1.60 binaries in the stellar subsample and 0.44 in the brown dwarf subsample. However after 
correcting for the limited sensitivity to such spurious close binaries with a large difference
in flux (using the observed Luminosity Function) the contamination drops to $\approx 0.79$ 
binaries in the stellar subsample and $\approx 0.17$ binaries in the brown dwarf subsample. 
Since these chance association binaries may include pairings with sources outside each 
subsample they are expected to include only 1.22 stars from the stellar subsample 
and 0.20 brown dwarfs from the brown dwarf subsample.
We observe 13 stars in binary systems, reducing to 11.78 after correcting for chance 
alignment. If the incidence of binarity in brown 
dwarfs is equal to that of stars then we would expect $3.22 \pm 1.79$ members of the 
brown dwarf sample to be in a binary system (assuming Poisson statistics), compared to 1 
which is observed. The expected chance contamination increases the expected number by
0.20 sources to $3.42 \pm 1.80$. We can therefore state that the incidence of binarity in 
brown dwarfs is therefore lower than in stars at the 1.34$-\sigma$ level of confidence.
The probability of this occurring by chance is only 9.0\% (using a one tail test).
A similar calculation for the closest populated bin only gives a 1.46$-\sigma$ result, based
on the detection of 8 stars located in binaries and no brown dwarf binaries. This
result has a 7.2\% probability of occurring by chance.

There is no obvious reason why the binarity fraction should be linked to the hydrogen burning
threshold at 0.075~M$_{\odot}$. Most of the stars detected in binaries by Gemini (see Table 3) 
have masses M$\sim 0.2$~M$_{\odot}$ (which is a typical stellar mass in the cluster) and none 
have masses M$< 0.13$~M$_{\odot}$. If we extend the brown dwarf subsample to include very low 
mass 
stars with masses up to 0.10~M$_{\odot}$ then the sizes of the subsamples becomes a little more 
equal: 190 stars and 71 very low mass stars and brown dwarfs. In these two bins the binary 
fraction in the less massive sample is then lower than the high mass sample with 1.75-$\sigma$ 
confidence, which has only a 4.0\% probability of occurring by chance.

\subsubsection{Subaru and HST/NICMOS data}
To investigate further the low incidence of binarity in brown dwarfs
a similar analysis was done with 3 additional datasets: the UKIRT data of LR00 and 
Lucas et al.(2001), the Subaru data of Kaifu et al.(2000) and the NICMOS results tabulated
in Luhman et al.(2000). Very few new subarcsecond binaries were clearly resolved in the 
UKIRT data. By contrast many additional binaries were resolved in a reinspection of the 
Subaru data, which had a slightly better full width half maximum (0.5'') than UKIRT and much 
less power in the wings of the instrumental profile. These additional detections (see Table 3) 
were often confirmed by less clear detections in the UKIRT data. 

These additional samples were chosen so as to be independent of the Gemini sample
and sensitive to both brown dwarf binaries and stars. The samples are:

\begin{itemize}
\item the part of the Subaru dataset located at $\theta>60''$ from $\theta_1$~Ori~C 
and excluding the Gemini survey region. This survey region (area 14.0 arcmin$^2$) produced 
a sample of 
312 stars and 57 brown dwarfs, the assignations being decided by source luminosity with 
the aid of the IJH band photometry of LR00 and the HK band photometry of Hillenbrand
\& Carpenter (2000). 10 stellar binary systems were resolved in this sample at $\theta<0.79''$
and 0 brown dwarf binaries. However, 2 of the closest stellar binaries are excluded
from our analysis since they were only marginally resolved by Subaru and required 
confirmation by the shallow HST/WFPC1 survey of Prosser et al.(1994), which was not
sensitive to brown dwarfs. These 2 binaries were not resolved in the published 
NICMOS/NIC3 photometry (0.4'' resolution). 

\item the part of the NICMOS dataset located at $\theta<60''$ from $\theta_1$~Ori~C.
The NICMOS data produced a sample of 159 stars and 43 brown dwarf candidates. In addition
7 sources of uncertain status were detected at 1.6$~\mu$m only  and these were excluded from 
the subsamples searched for binarity. The assignation of each source to a subsample was 
aided by the K band photometry of Hillenbrand \& Carpenter (2000). Somewhat surprisingly no 
binaries at all were detected in this sample with $\theta<0.79''$. 2 stellar binaries happen
to straddle the 60$^{``}$ boundary but these were included in the Subaru sample. 
\end{itemize}

\begin{figure*}
\begin{center}
\begin{picture}(200,400)

\put(0,0){\includegraphics{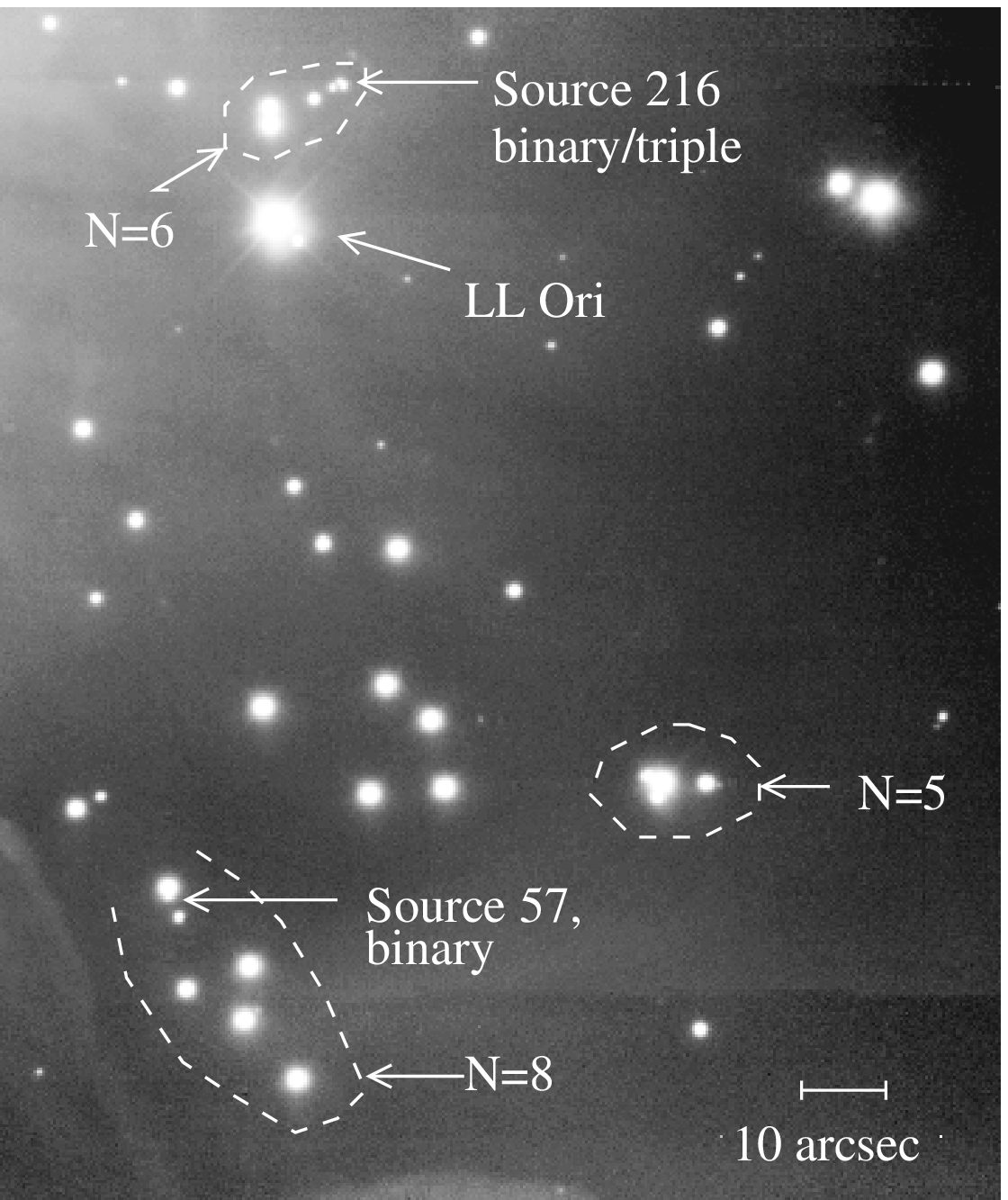}}

\end{picture}
\end{center}
Figure 13. H band image of part of Field 2. Some possible small N associations
are indicated with dashed lines, though the boundaries of these are often unclear.
The location of 2 brown dwarf candidates located in apparent subarcsecond binaries
are indicated, though these are not clearly separated from their brighter companions 
in this large scale view.
\end{figure*}

This absence of binaries in the NICMOS sample is surprising given the relatively high 
incidence of binarity measured further from the cluster centre in the Subaru and Gemini 
datasets. It might be due to the higher stellar density at $\theta<60''$, which leads to a 
greater frequency of close encounters capable of disrupting wide binaries with separations 
$> 200$~AU. Scally \& Clarke (2001) calculated that such close encounters are rare in
typical, less dense parts of the cluster. This explanation is supported by the observation
that closer, more tightly bound stellar binaries $\theta \le 0.35''$ do exist
in this central region of the cluster, eg. Prosser et al.(1994), Petr et al.(1998). 

The surface density in the Subaru sample region is $\Sigma=7.3 \times 10^{-3}$arcsec$^{-2}$.
It is sufficiently uniform (variations of $\pm 25\%$ seen in different 1~arcmin$^2$ areas
of the region) to be assumed uniform for our calculations of chance contamination.
After accounting for selection effects and pairing with objects outside each subsample
the expected number of objects found in a binary by
chance alignment is 5.0 in the stellar subsample and 0.47 in the brown dwarf subsample.
Given that 16 stars are observed in binaries in the stellar subsample the expected number
of genuine binary members in the brown dwarf subsample would be $2.01 \pm 1.42$ if the 
incidence of binarity and the detectability of binaries were the same as in the stellar 
subsample. However the Subaru brown dwarf subsample includes many sources near the
sensitivity limit of that relatively shallow survey, so the sensitivity to brown dwarf 
binaries is only $\approx 75\%$ of the sensitivity in the stellar subsample (due to
a reduction of a factor of approximately 2 in the sensitivity to fainter companions).
Hence the expected number of genuine binary members is only $1.51 \pm 1.23$ if brown dwarf
binaries are as common as stellar binaries. The expected chance alignment (an extra 0.47 
sources) can be neglected in this subsample: it is known to be zero since no brown dwarf 
binaries are observed in the Subaru data. This indicates that
the incidence of binarity is lower than in stars at the 1.23$-\sigma$ level, which
has a probability of 10.9\% (one tail test). By itself this is a very low level of 
significance but taken together with the 9.0\% probability from the independent Gemini 
sample it provides a fairly strong result with 99.0\% confidence (one tail test), 
or 96.1\% confidence in a 2 tail test.

\begin{figure*}
\begin{center}
\begin{picture}(200,210)

\put(0,0){\includegraphics{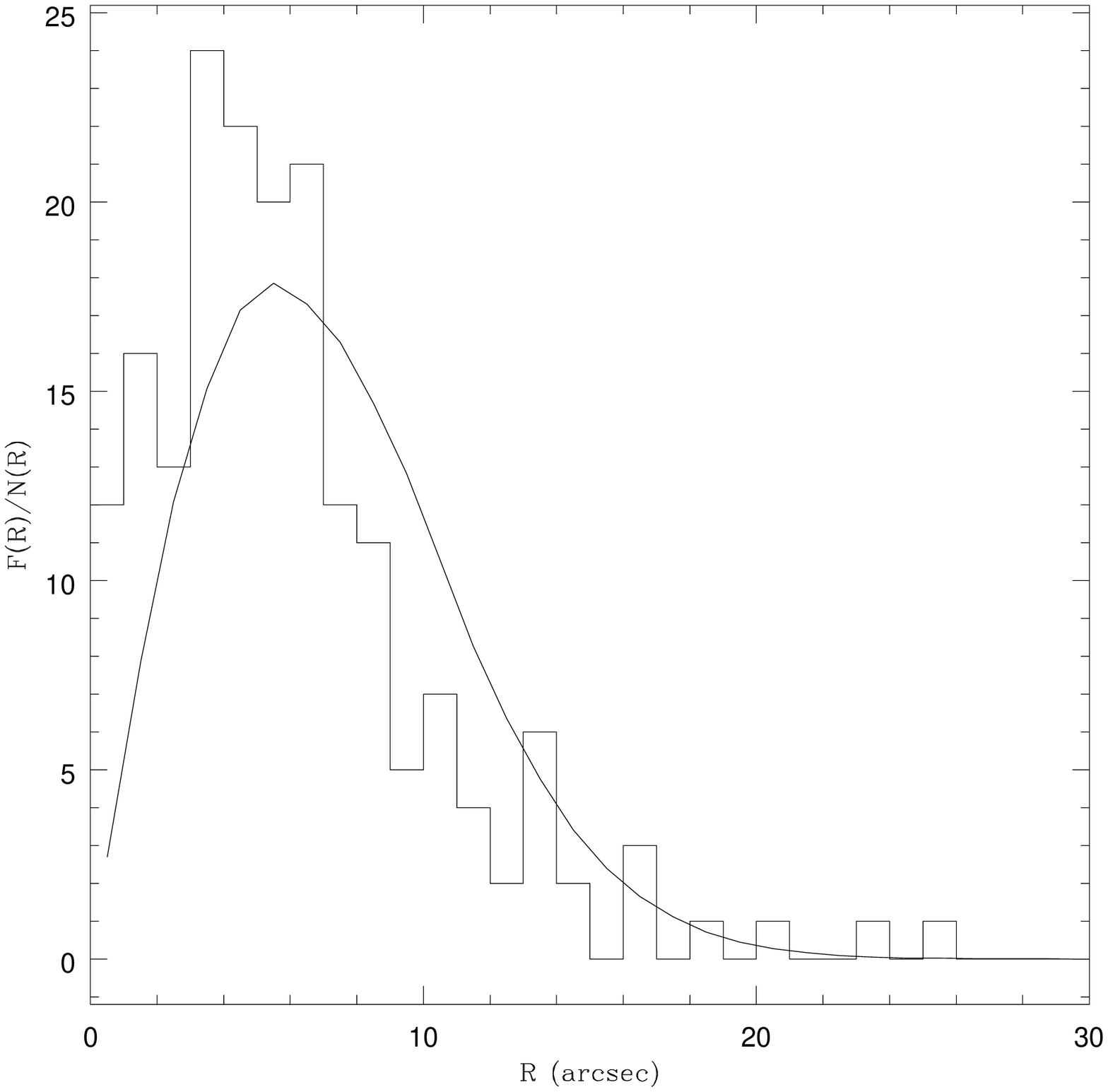}}

\put(0,0){\includegraphics{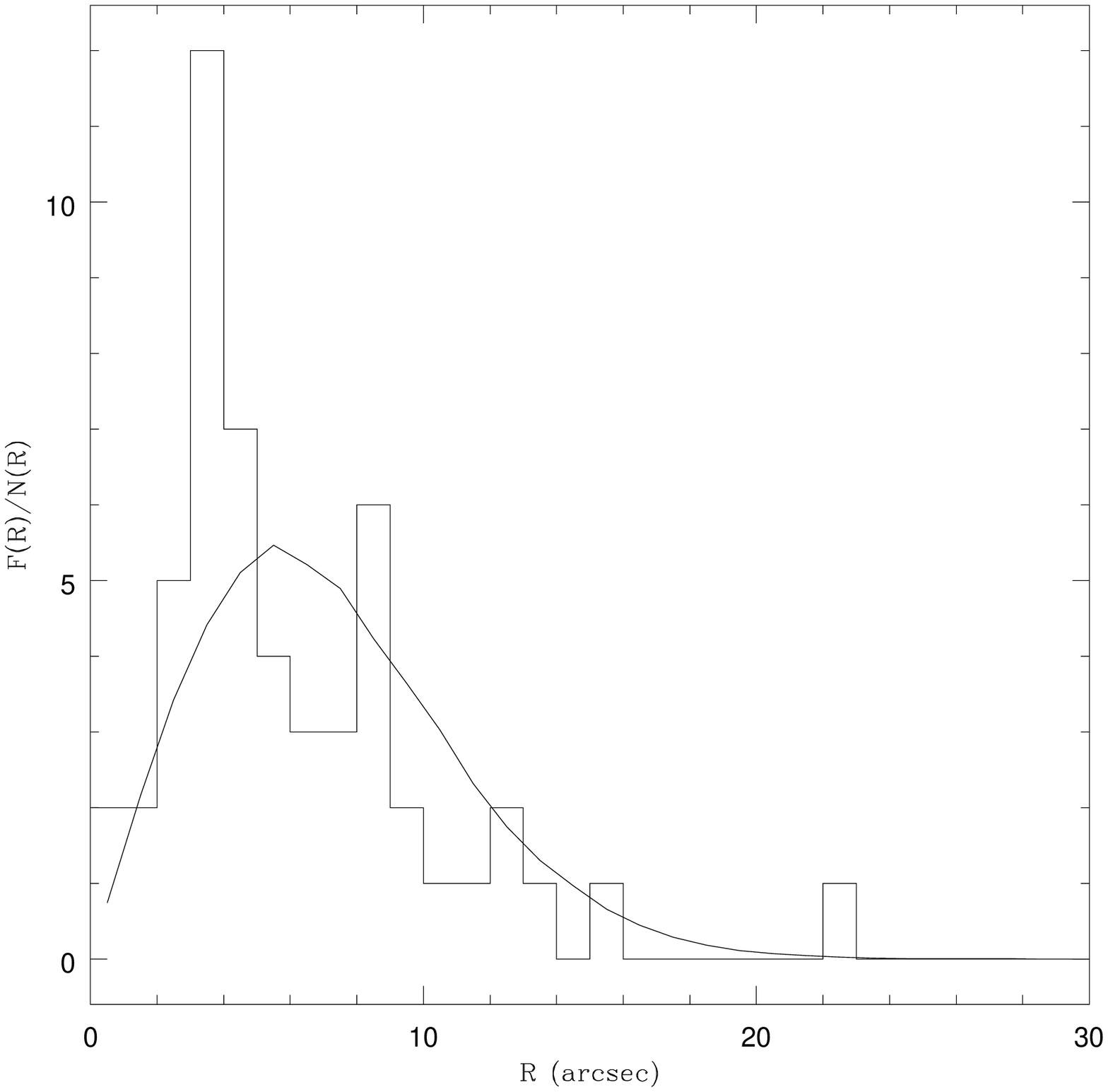}}

\end{picture}
\end{center}
Figure 14. Nearest neighbour functions for the Gemini sources, correlating
the stellar and substellar samples with all detected sources. (left) stars. 
(right) brown dwarfs. The curves represent the expected function for a random distribution.
A slight excess over the random distribution is seen for stars at separations of 
3-7 arcsec, supporting the impression of some small N groups. The brown dwarf histogram 
has more scatter but shows an excess of nearby companions at separations of 2-5 arcsec.
These excesses appear to indicate the existence of small-N associations.
\end{figure*}

Again, the significance of the result from the Subaru data is increased if the brown dwarf
subsample is enlarged by including very low mass stars with M$<0.1$M$_{\odot}$. In this 
case there are 293 sources in the stellar subsample and 76 in the low mass object subsample,
which still contains no binary members. Performing the same type of calculation (and retaining
the 75\% sensitivity factor for the low mass susbsample) the expected number of binary 
members in the low mass subsample is $2.24 \pm 1.50$. The null detection of low mass binary
members therefore has a 6.7\% probability of occurring by chance. Combining this result with 
the Gemini result gives 99.7\% confidence (one tail test) that the binary fraction is lower 
among sources with M$<0.1$M$_{\odot}$, or 98.9\% in a two tail test.

\subsubsection{Discussion: binaries and associations}
The apparent paucity of low mass sources in binary systems is interesting because it provides
support for the ejected stellar embryo hypothesis of brown dwarf formation. 
The SPH simulation of Bate et al.(2003) predicts that $\sim 75\%$ of proto-brown dwarfs 
form via disc fragmentation and 25\% form via fragmentation of filaments but in both cases 
it is the subsequent dynamical interaction and ejection from the gas reservoir that 
prevents the brown dwarfs from reaching stellar mass. In this scenario only a very small 
fraction of brown dwarfs would be expected to be binary, perhaps in cases where a tight 
binary proto-brown dwarf is ejected as a single unit from a high order multiple system 
(see Sterzik \& Durisen 1998).
As discussed earlier wide binaries ($>300$~AU in the cluster centre) are thought to
have been disrupted by close stellar encounters. However, in the less dense outer region
of the cluster observed here only a small fraction of sources are likely to have had an 
encounter at $<400$~AU distance with a star that did not form in the same group 
(see Scally \& Clarke 2001) so the observed incidence of binarity is (to first order)
primordial. 

There have in recent years been several investigations of the binary fraction in brown 
dwarfs in the local field and in older galactic clusters such as the Pleiades 
(eg. Pinfield et al. 2003; Reiners \& Basri 2005; Close et al. 2003; Burgasser et al. 2003;
Gizis et al. 2003; Martin et al. 2003; Reid et al. 2001; Koerner et al. 1999; Potter et al.
2002). These surveys have established that wide brown dwarf-brown dwarf binaries ($>20$~AU) 
are extremely rare and the overall distribution of orbital semi-major axes peaks at 
much smaller radii than in stars. Wide binaries consisting of a solar type star and a brown dwarf
may be more common however (eg. Gizis et al. 2001; Metchev \& Hillenbrand 2004; Volk et al.2003; 
McCaughrean et al. 2004). Earlier surveys of stellar binarity (eg. Duquennoy \& Mayor 1991; 
Fischer \& Marcy 1992) showed that the binary fraction decreases slightly  
as a function of the mass of the primary from $\sim 57\%$ in solar type stars to $\sim 35\%$ 
in early M types. At separations $>1-2$~AU the direct imaging surveys listed above found a 
binary fraction of $\sim 15\%$ in late M and L types, and perhaps even less in T types. 
However, Pinfield et al.(2003) detected a much higher incidence of binarity in Pleiades
brown dwarfs (30-50\%) by detecting binaries photometrically. This suggested that either
there should be many unresolved binaries in the local field or the Pleiades population
is unusual. The former conclusion is given some support by the recently reported detection of 
a few close substellar spectroscopic binaries (Reiners \& Basri 2005) and the previous 
detection of one such object by Basri \& Martin (1999).

The low incidence of wide binaries containing brown dwarfs in Orion at separations 
$r \ge 150$~AU is unsurprising given the total absence of such objects in the local field 
population. However the significance of this result is that it holds true for this
primordial population and is therefore likely to be a product of the star formation
process. 
Note that in the samples descibed here brown dwarfs can only be detected in
binaries where the primary is a low mass star or brown dwarf, owing to the limited dynamic
range at separations $\theta < 0.79$ arcsec.
It is perhaps more surprising that at least 1 brown dwarf (Source 216) is detected in a probable
wide star/brown dwarf binary with a luminosity contrast of $\sim~1$ magnitude.
Neither component of the system shows any sign of an (H-K) excess.
Only one wide brown dwarf binary pair has been reported in the 
literature to date, an $r=240$~AU pair in the Chamaeleon star formation region (Luhman 2004),
though there is no firm evidence that this is a bound system. In the case of the Orion binary 
brown dwarf candidate, Source 216, it may be significant that it is located in an apparent 
triple system around which is a small N subgroup of perhaps 6 to 10 sources (see Figure 13). 
The tertiary component is 1.2$\arcsec$ distant from the nearest component of the binary. 
This subgroup is the active region
of recent star formation associated with LL Ori (see Bally, O'Dell \& McCaughrean 1999),
in which several stellar outflows are apparent, confirming the extreme youth of the group.

Another brown dwarf candidate from the Gemini dataset, Source 57, 
is also observed in a possible binary with a 0.93 arcsec separation. It is also a member of an 
apparent small-N subgroup in Field 2, with about 8 members. The subgroup group containing 
Source 57 contains a second brown dwarf 
(Source 40) located only 2 arcsec from a bright star, and there are other instances of 
possible brown dwarf binaries in Field 2 in the $\theta=1-4^{``}$ separation range. These 
apparent binaries and small-N groupings appear more significant in a 
visual inspection of Field 2 (see Figure 13) than they would in other parts of the survey 
region due to the slightly lower source density there.

The reality of the small N groupings and $\theta=1-4$ arcsec star $+$ brown dwarf binaries 
is given statistical support by the two point correlation
functions in Figure 12 and the nearest neighbour distributions in Figure 14. 
At $\theta<6$ arcsec Figure 12 shows that $w(\theta)$ is generally slightly above 
zero in both the stellar and substellar populations, suggesting some tendency for sources to 
form in small groups. The nearest neighbour distributions plotted in Figure 14 support this 
impression, particularly for the stellar population where the histogram consistently lies 
above the solid curve representing the expected number of nearest neighbours per bin
for a random distribution at separations of 3 to 7 arcsec (1300 to 3000~AU).
The probability of such a chance excursion occurring in these 4 bins is only 
0.12\% (two tail test). The nearest neighbour distribution for brown dwarfs shows more 
scatter about the curve for a random distribution due to the smaller sample size. 
However there is an excess in the distribution at $\theta=2-5$ arcsec, which 
has a 0.07\% probability of occurring by chance in these 3 bins (two tail test) and
only a $\sim 0.25\%$ probability of occurring somewhere in the plotted distribution.

It therefore seems very probable that a significant fraction of Trapezium brown dwarfs and 
stars have formed in small-N associations. We note that the fraction of sources presently 
appearing to be in such associations is small.
The simulations of Scally \& Clarke (2002) indicate that subclustering is removed within a few 
Myr in dense star forming regions such as the Trapezium Cluster. It is therefore unsurprising 
that the apparent associations are only intact and observable in the sparse outer regions of the 
cluster that are observed here. Bate et al.(1998) found
no sign of such sub clustering in the data of McCaughrean et al.(1996) and Prosser et al.(1994)
for the central regions.
Their plot of the MSDC function for the large area survey of Jones \& Walker (1988) did
reveal a slightly higher incidence of companions at separations of a few arcsec which might be 
interpreted as evidence of small-N clustering. However, the shallow optical survey of 
Jones \& Walker was not optimal for detection of clusters with low mass members.
Bate et al.(1998) emphasised the difficulties in interpreting the larger scale slope:
MSDC$(\theta) \propto \theta^{-0.16 \pm 0.01}$. 

Following the hypothesis of Reipurth \& Clarke (2001) we suggest that 
these associations may be dynamically unstable. It is possible that some members of the 
apparent associations are in the process of being ejected. However, the observed associations 
appear to suffer little extinction and may
be too mature for future ejections to aid in the production of brown dwarfs.
It is of course very possible that the 2 brown dwarfs in apparent sub-arcsecond binaries are 
chance associations caused by projection effects (most likely projection of cluster members 
rather than background stars for these fairly bright sources). The probability of a chance 
association is $\approx 17\%$ for Source 216 and close to $\approx 30\%$ for Source 57. The 
main point is that there is a statistically significant excess of close neighbours at 
separations $\theta<$5 arcsec in the nearest neighbour distribution. Kinematic measurements will 
be needed to explore the nature of these apparent associations and wide binaries.

\section{Discussion of Nebulae, Trails and Filaments}
\subsection{Trails and star forming filaments?}

In this section we present a speculative discussion of the luminous trails 
and filaments with which some sources seem to be associated. In Lucas, Roche 
\& Riddick (2003) we noted that a minority of the low luminosity point sources in
Field 1 appear to have short ($\sim 4$~arcsec) trails of continuum radiation extending 
away from them, usually in the direction of the cluster centre. This orientation
rules out identifying them as photoevaporating 'proplyds', the mainly neutral 
cometary nebulae pointing away from the cluster centre observed by eg. O'Dell \&
Wen (1994) in Orion and Hester et al.(1996) in M16. We speculated that they might
be associated with source motion away from the cluster core. The Trapezium cluster is 
pervaded
by highly structured nebulosity, so it would be preferable to conduct a statistical 
analysis to determine whether the apparent association of the trails with a few point
sources is genuine. However, on inspection of the data we were unable to find 
a useful definition of which patches of nebulosity with no point source association 
might constitute a trail. In consequence we limit this discussion to reporting 
the following observational features which may be of interest.

\begin{figure*}
\begin{center}
\begin{picture}(200,240)

\put(0,0){\includegraphics{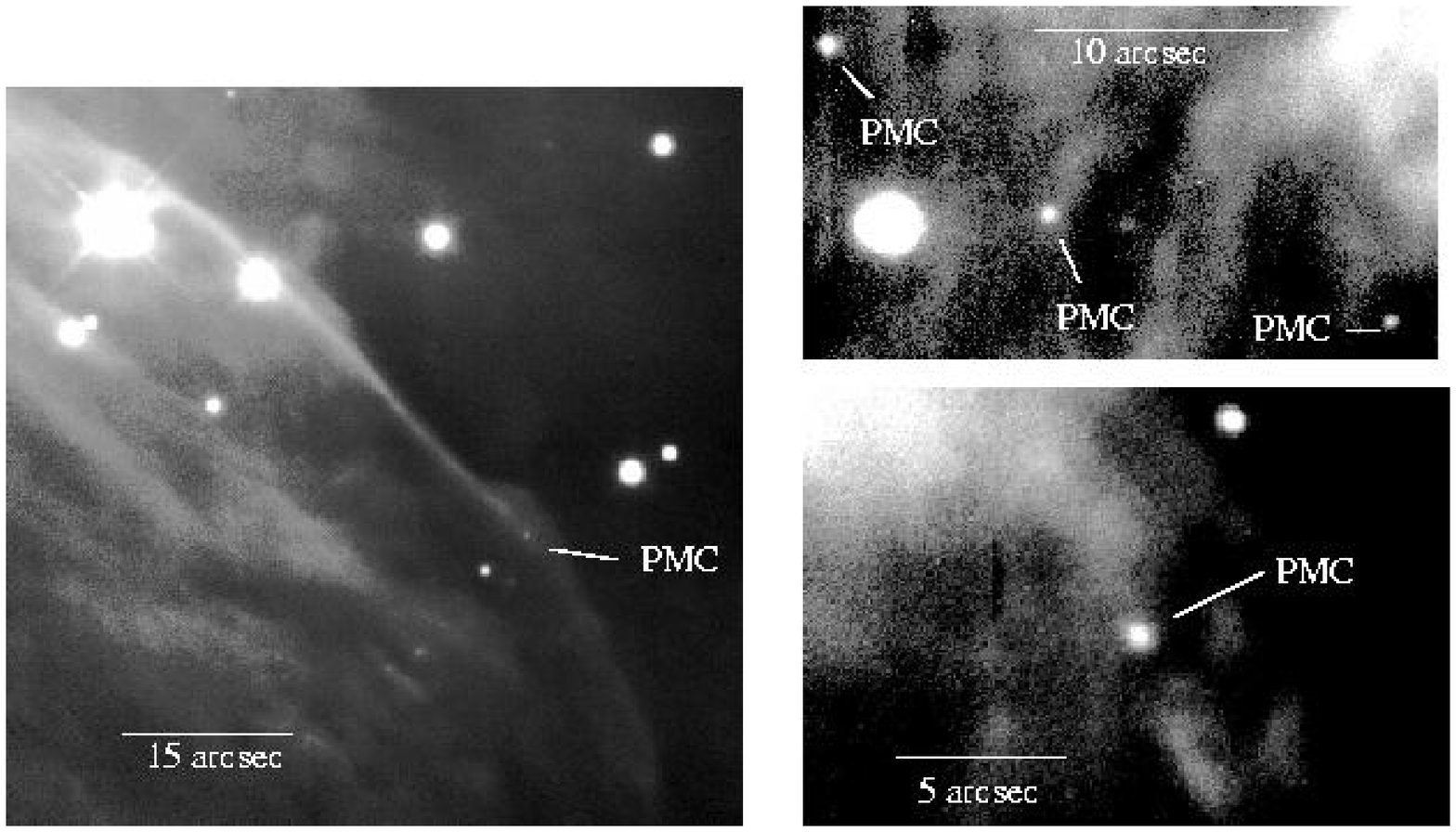}}

\end{picture}
\end{center}
Figure 15. 'Trails' associated with 3 PMCs. The very long trail in the left panel
appears similar to the filaments of nebulosity shown in Figure 16, except that a PMC
(Source 64) lies at one end. The upper right panel shows 3 PMCs in a small area of Field 1.
The central PMC (Source 109) lies at one end of a short streak of nebulosity and also 
exhibits a K band excess of 0.3 magnitudes.
\end{figure*}

\begin{figure*}
\begin{center}
\begin{picture}(200,340)

\put(0,0){\includegraphics{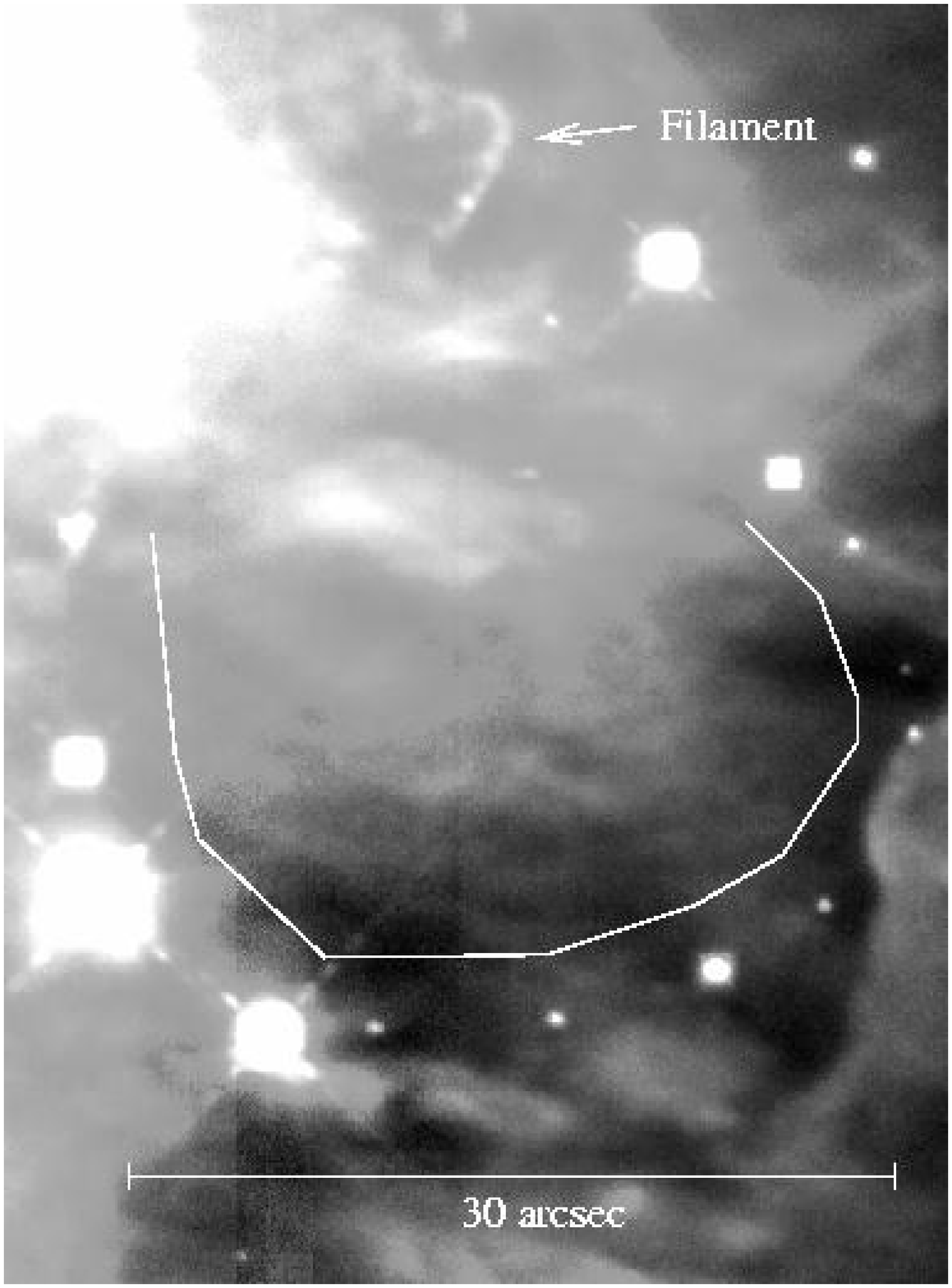}}

\put(0,0){\includegraphics{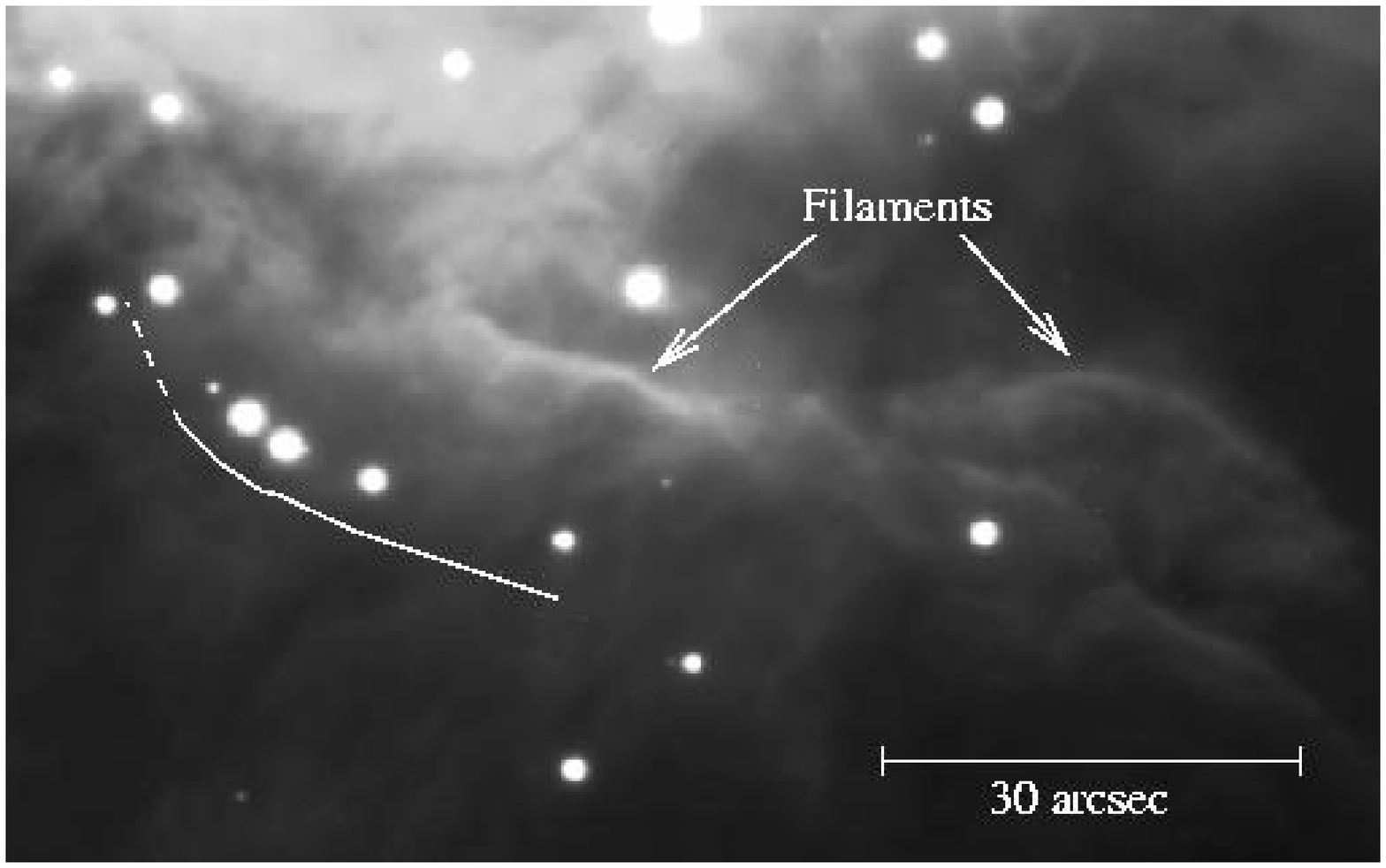}}

\end{picture}
\end{center}
Figure 16. Filaments and associations. Possible associations are illustrated with curved 
white lines (see text). In the upper part of the left panel a short curved filament is 
seen. This filament contains a single point source, Source 329 (see text) and may be 
dominated by H$_2$ emission. The right panel shows filaments within the Orion Bar.
\end{figure*}

\begin{itemize}
\item Trails. A total of 8 sources in Table 2 appear to lie at one end of
a short and narrow 'trail' of nebulosity (see Figure 15). There is no sign of the
V-shaped bow shocks which might be expected if these are H$_2$ outflow features,
and in some cases the trails are as bright at H band as at K band.
Of these, 3/8 are PMCs (Source 64=080-646, Source 109=199-617 and Source 173=092-532), 
3/8 are brown dwarf candidates (Source 103=179-621, Source 370=044-219 and Source 
128=157-602) and 2/8 are of stellar mass (Source 46=139-701 and Source 252=105-417). 
The trails of the 3 PMCs 
all point in the approximate direction of the cluster centre (within 30$^{\circ}$) and 
3 of the more luminous sources (Sources 103, 128 and 252) share this 
characteristic. One of the two with a different direction, Source 370, has a trail which 
appears narrow and linear in our UKIRT data from LR00 at H band, but the Gemini data 
reveals a fainter patch of more diffuse nebulosity associated with the source which make 
the designation as a trail questionable. The other, Source 46, is a J mag=20 stellar mass
source with 
high extinction (A$_V=40$) that has a 4 arcsec (2000~AU) trail which is only visible at J 
band, being swamped by the wings of the stellar profile at H and K. Similar faint
trails would go undetected in most stellar mass sources with lower extinction.
If any of these trails are associated with source motion they might either be 'light
pipes' of relatively low extinction internally illuminated by the associated source,
or local density enhancements which are externally illuminated by the ambient radiation 
field. 

\item Filaments. Several curved filaments are observed in this survey, some with
no associated point source. Examples are illustrated in Figures 15 and 16.
These filaments are not obviously associated with any high luminosity source that might
power an outflow. We speculate that some of them might be dense star forming 
filaments of the type produced in SPH simulations of star formation in a turbulent gas 
cloud (eg. Bate, Bonnell \& Bromm 2003). The filaments in the left panel of Figure 15 
and the right panel of 
Figure 16 appear bright at J, H and K, while the one in the left panel of Figure 16
is bright only at K band, suggesting that it may be H$_2$ emission. This latter
filament is a highly curved and $\sim 10$ arcsec long. It runs through an unresolved 
object, Source 329 (=093-255). The filament includes several 
marginally resolved knots or point sources seen only at K band, which may be knots
of H$_2$ emission. Source 329 might also be merely a gaseous knot but 
it is included in Table 2 as a stellar source since it is detected at J, H and K, though 
it was not sufficiently well resolved for photometry at H band. The filament shown
in the left panel of Figure 15 is the 'trail' associated with Source 64 that was described 
above. However its great length ($\sim~1$~arcminute$\approx 25000$~AU) and curvature 
suggest that it is perhaps more likely to be a product of the turbulent star formation 
process than gas associated with a single very low mass source. High resolution observations
in the millimetre, submillimetre or far infrared wavebands could test whether any of
these filaments are in fact composed of cold and very dense gas. The submm data shown in 
Figures 1 and 7 reveals that the bright Orion Bar, which appears as a giant, externally
excited luminous filament in the infrared, has an optical density approaching A$_V=100$ 
in places, so the existence of smaller high density structures that were unresolved
by SCUBA is possible.

\item Small-N associations. We noted in the previous section that there are several 
apparent small-N associations which appear at a statistically significant level
in the nearest neighbour function. In the context of the possible observation of
dense star forming filaments we note that two suggestive string-like groups of stars 
are seen in this survey (see Figure 16), both of which lie close to filamentary 
nebulosities. It is likely that these groupings are simply a chance temporary alignment 
of sources whose origins are unconnected. However,
it is possible that these apparent associations are relics of a filamentary 
mode of star formation, being cases where the filament has been dispersed and is no longer
observable. Most Trapezium cluster members have a random velocity dispersion of 
$\sim 3.5$~kms$^{-1}$ in the plane of the sky (Jones \& Walker 1988). Hence, such apparent 
associations cannot survive more than a few thousand years if their members share this random 
velocity dispersion. However if sources formed in a filament retain a common proper motion 
then the spatial association might remain detectable for a few $\times 10^4$~yr, which
is perhaps the youngest conceivable age for these sources.
On a larger scale it is widely accepted that many of the YSOs in the nearby 
Taurus-Auriga cloud complex formed in filaments perpendicular to the local magnetic field
which are no longer directly observable (see Menard \& Duchene 2004 and references therein). 
The existence of these possible associations might be tested by spectroscopy, searching
for indicators of extreme youth, or a high precision proper motion survey. 
\end{itemize}

\begin{figure*}
\begin{center}
\begin{picture}(200,240)

\put(0,0){\includegraphics{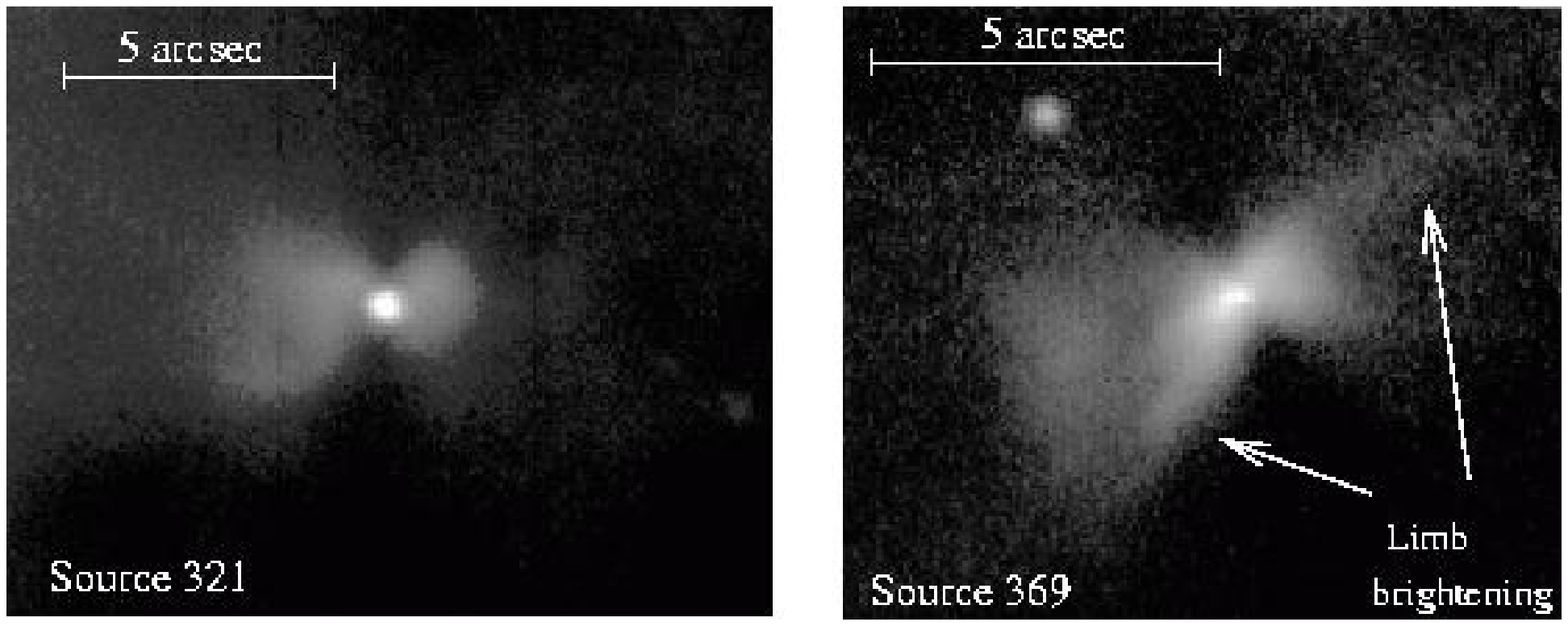}}

\end{picture}
\end{center}
Figure 17. K band images of Bipolar Nebulae. (left) Source 321 (057-305), (right) 
Source 369 (013-220). North is up and east is to the left. Source 321 is also visible 
at K, H and J in Figure 11. 
\end{figure*}

\subsection{Bipolar Nebulae}

We report the discovery of two bipolar nebulae with bright central sources in Field 3 
(see Figure 17) which are included in Table 2. 
Both these sources (Source 321 (057-305) and Source 369 (013-220)) have previously been 
imaged in the UKIRT data of LR00 and Source 321 was also imaged by both Hillenbrand \& 
Cerpenter (2000) and Muench et al.(2002). In those surveys these nebulae went unremarked: 
in the UKIRT data both appear merely as an elongated nebula associated with the central 
source. However, the high resolution of the 
Gemini K band data in Field 3 (0.36 arcsec) reveals the hourglass-like indentations in the 
nebulosity on either side of the central source that is commonly observed in equatorially
condensed nebulae whose axis of rotational symmetry lies close to the plane of the sky.
These systems are likely to be Class I YSOs similar to those observed in nearby low mass
star formation regions.

Source 321 is an unresolved source seen at J, H and K at the centre of the surrounding 
nebula. At J and H band the contrast with the surrounding nebula is low so it is possible 
that it is merely a well defined scattering peak. At K band the point source is much brighter
than the surrounding nebula, indicating that a protostar is being observed directly
(perhaps with a contribution from hot circumstellar dust close to the star).
Source 321 appears to be similar to the famous Serpens Reflection Nebula 
(SRN, also known as Serpens SVS2), in which the central source is visible at near infared
wavelengths despite the presence of relatively high extinction in the equatorial plane.
Most YSOs with a bipolar nebulae have a dark central lane which completely obscures
the central source, even in more evolved Class II systems such as HK Tau b (Stapelfeldt 
et al. 1998).
It is unclear why the central source is visible in some bipolar systems. It is possible that
the circumstellar matter is disturbed by an outflow from the YSO, allowing light to
shine through a hole in the matter distribution. Another possibility is that these systems
have unusually small accretion discs surrounded by a circumstellar envelope which has
only a weak vertical density gradient (perhaps due to a low systemic angular momentum).
A small disc ($r \ltsimeq 100$~AU) in vertical hydrostatic equilibrium would have too 
small a scale height to block light from the central source unless the disc axis is 
almost exactly in the plane of the sky.
 
By contrast, Source 369 is marginally resolved at all 3 wavelengths, indicating that
this probably is a scattering peak and the protostar is obscured from view. The cavity
walls appear to be limb brightened (a well known geometric effect) but interestingly
this is obvious only for the north western and south eastern walls. 
This phenomenon has been explained using a tilted disc model (Gledhill 1991) 
but might also be due to shadowing of the fainter cavity walls by irregularly distributed 
matter within the bipolar cavity.

\section{Conclusions}

This very deep survey of the outer parts of the Trapezium Cluster
with Gemini South/Flamingos has allowed us to produce
an extinction limited sample (A$_V<5$) complete to a mass of approximately 
0.005M$_{\odot}$ (5~M$_{Jup}$), for sources with an age of 1~Myr. 
The initial mass function appears to drop by a factor of $\sim 2$ at the 
deuterium burning threshold, although this may be due to a smooth decline
in the mass function at planetary masses which appears abrupt because 
of the small sample size.

An upper limit of 13\% is placed on the number fraction of planetary mass 
objects (PMOs) in the range 3-13~M$_{Jup}$, based on the assumption that all
candidates are genuine 1~Myr old cluster members. However, background 
contamination appears to rise steeply at luminosities corresponding to 
M$<5$~M$_{Jup}$, indicating that the true number fraction of PMOs is probably 
$<10\%$. Spectroscopic follow up of a large sample of candidates will be 
required to establish the true number fraction and test whether there is a 
cut off as predicted by the theory of opacity limited fragmentation. 
Our attempt to test for (H-K) excesses in substellar candidates produced
inconclusive results owing to uncertainties in the intrinsic colours of
young brown dwarf photospheres and the limited photometric precision of the 
dataset. However, in this survey region we find no sign of the proplyds with 
bluer (H-K) colours than the isochrone which were detected by Muench et al.
(2001) nearer the cluster centre. This may be due to a slightly greater age 
for sources in the outer part of the cluster.

Analysis of the spatial distribution of stars and brown dwarf candidates 
suggests that brown dwarfs and very low mass stars (M$<0.1$M$_{\odot}$) are less 
likely than more massive stars to have wide ($>150$~AU) binary companions. This 
result has modest statistical significance in our data (96\% in a one tail test)
but is supported at 93\% confidence by analysis of an completely independent sample 
taken from the Subaru data. Intriguingly, there is a 
statistically very significant excess of both stars and brown dwarfs with small
separations from each other ($<6$ arcsec or 2600~AU). This appears to be due
to the presence of small N subgroups, which are likely to be dynamically unstable 
in the long term. These results might be interpreted as providing support for
the 'ejected stellar embryo' hypothesis for brown dwarf formation 
(Reipurth \& Clarke 2001). 

\section{Acknowledgements}
 
        We are very grateful to the Gemini staff who made the observations
and to the University of Florida team who provided Flamingos I. This paper
is based on observations obtained in programme GS-2001B-Q-11 at the Gemini
Observatory, which is operated by the Association of Universities for Research
is Astronomy, Inc., under a cooperative agreement with the NSF on behalf
of the Gemini partnership: the National Science Foundation (USA), the 
Particle Physics and Astronomy Research Council (UK), the National Research 
Council (Canada), CONICYT (Chile), the Australian Research Council (Australia),
CNPq (Brazil) and CONICET (Argentina). We also wish to thank Professor Norio
Kaifu for permitting use of the Subaru Telescope first light data and 
Dr Doug Johnstone of the Canadian National Research Council for providing
the submm data used to illustrate the background extinction.
PWL is supported by PPARC via an Advanced Fellowship at the University of 
Hertfordshire. Finally we thanks the anonymous referee for a helpful report.

\section{References}
Baraffe I., Chabrier G., Allard F., Hauschildt P.H. 1998, A\&A, 337, 403\\
Baraffe I., Chabrier G., Allard F., Hauschildt P.H. 2002, A\&A, 382, 563\\
Basri G., Martin E.L. 1999 AJ, 118, 2460\\
Bate M.R., Clarke C.J., McCaughrean M.J. 1998, MNRAS 297, 1163\\
Bate M.R., Bonnell I.A. Bromm V. 2003, MNRAS 339, 577\\
Béjar V.J.S., Martín E.L., Zapatero Osorio, M.R. Rebolo R., Barrado y Navascués D., 
  Bailer-Jones C.A.L., Mundt R., Baraffe I., Chabrier C., Allard F. 2001, ApJ 556, 830\\
Boyd D.F.A., Whitworth A.P. 2005, A\&A 430, 1059\\
Burgasser A.J., Kirkpatrick J.D., Reid I.N., Brown M.E., Miskey C.L.,
  Gizis J.E. 2003, ApJ 586, 512\\
Burgasser A.J. 2004, ApJS 155, 191\\ 
Cardelli J.A., Clayton G.C., Mathis J.S., 1989, ApJ, 345,245\\
Carpenter J.M., Hillenbrand L.A., Skrutskie M.F. 2001, AJ, 121, 316\\
Carpenter J.M. 2003, on-line at http://www.astro.caltech.edu/~jmc/2mass/v3/transformations/\\
Chabrier G., Baraffe I., Allard F., Hauschildt P. 2000, ApJ 542, 464\\ 
Close L.M., Siegler N., Freed M., Biller B. 2003, ApJ 587, 407\\
Cohen J.G., Frogel J.A., Persson S.E., Elias J.H. 1981, ApJ 249, 500\\
Dahn C.C., Harris H.C., Vrba F.J., Guetter H.H., Canzian B., Henden A.A., Levine S.E.,
  Luginbuhl C.B., +10 authors, 2002, AJ, 124, 1170\\
D'Antona F., Mazzitelli I. 1997, Mem. Soc. Astron. Ital. 68, 807\\
Elston R., 1998, Proc. SPIE 3354, 404\\
Gizis J.E., Kirkpatrick J.D., Burgasser A., Reid I.N., Monet D.G., Liebert J., 
  Wilson J.C. 2001, ApJ 551, L163\\
Gizis J.E., Reid I.N., Knapp G.R., Liebert J., Kirkpatrick J.D., Koerner D.W.,
  Burgasser A.J. 2003, AJ 125, 3302\\
de Grijs, R.; Johnson, R. A.; Gilmore, G. F.; Frayn, C. M. 2002a, 331, 228\\
de Grijs, R.; Gilmore, G. F.; Johnson, R. A.; Mackey, A. D. 2002b, 331, 245\\
de Grijs, R.; Gilmore, G. F.; Mackey, A. D.; Wilkinson, M. I.; Beaulieu, S. F.; 
  Johnson, R. A.; Santiago, B. X. 2002c, MNRAS 337, 597\\
Hillenbrand L.A. 1997, AJ 113,1733\\
Hillenbrand L.A., Carpenter J.M. 2000, ApJ 540, 236 (HC)\\
Hillenbrand L.A. \& White R., 2004, ApJ 604, 741\\
Johnstone D., Bally J. 1999, ApJ 510, L49\\
Jones B.F., Walker M.F. 1988, AJ 95, 1755\\
Kaas, A. 1999, AJ 118, 558\\
Kaifu N., Ususda T., Hayashi S.S., Itoh Y., Akiyama M., Yamashita T.,
  Nakajima Y., Tamura M. + 85 authors, 2000, PASJ 52, 1\\
Knapp G.R., Leggett S.K., Fan X., Marley M.S., Geballe T.R., Golimowski D.A.,
  Finkbeiner D., Gunn J.E., + 21 coauthors 2004, 127, 3353\\
Koerner D.W., Jensen E.L.N., Cruz K., Guild T.B., Gultekin K. 1999, ApJ 526, L25\\
Kroupa P., Petr M.G., \& McCaughrean M.J. 1999, New Astronomy 4, 495\\
Kroupa P., Bouvier J. 2003, MNRAS 346, 369\\
Lada C.J., Muench A.A., Lada E.A., Alves J.F. 2004, AJ, 128, 1254\\
Leggett S.K., Allard F., Geballe T.R., Hauschildt P.H., Schweitzer A. 
  2001, ApJ 548, 908\\
Liu, M. C., Najita, J., Tokunaga, A. T. 2003, ApJ, 585, 372\\
Low C., Lynden-Bell D., 1976, MNRAS 176, 367\\
Lucas P.W., Roche P.F. 2000, MNRAS 314, 858\\
Lucas P.W., Roche P.F., Allard F., Hauschildt P.H. 2001, MNRAS 326, 695\\
Lucas P.W., Roche P.F. \& Riddick F.C. 2003, in proc IAU Symposium 211, 'Brown Dwarfs',
  p63, ed. E.L. Martin, ASP, San Francisco\\
Luhman K.L., Rieke G.H., Young E.T., Cotera A.S., Chen H., Rieke M.J.,
  Schneider G., Thompson R.I. 2000 ApJ 540, 1016\\
Luhman K.L. 2004, ApJ 614, 398\\
Martin E.L., Brandner W., Bouvier J., Luhman K.L., Stauffer J., Basri G.,
  Zapatero Osorio M.R., Barrado y Navascues D. 2000, ApJ 543, 299\\
Masunaga H., Inutsuka S.-I. 1999, ApJ 510, 822\\
McCaughrean M., Zinnecker H., Rayner J., Stauffer J. 1995, in ``The Bottom of
  the Main Sequence and Beyond'', ed. C. Tinney, p209, (Berlin: Springer)
McCaughrean M.J., Close L.M., Scholz R.-D., Lenzen R., Biller B., Brandner W.,
  Hartung M., Lodieu N. 2004, A\&A, 413, 1029\\
Menard F., Duchene G. 2004, A\&A 425, 973\\
Metchev S., Hillenbrand L. 2004, Mem. Soc. Astron. Ital., 73, 23\\
Mohanty S., Basri G., Jayawardhana R., Allard F., Hauschildt P., Ardila D.,
  2004, ApJ 609, 854\\
Mohanty S., Jayawardhana R., Basri G. 2004, ApJ 609, 885\\
Mohanty S., Jayawardhana R., Basri G. 2005, ApJ (in press), astro-ph/0502155\\
Muench A.A., Alves J., Lada C.J., Lada E.A. 2001, ApJ 558, L51\\
Muench A.A., Lada E.A., Lada C.J., Alves, J. 2002, ApJ 573, 366\\
Muench A.A., Lada E.A., Lada C.J., Elston R.J., Alves J.F., Horrobin M., 
  Huard T.H., Levine J.L., Raines S.N., Roman-Zuniga C. 2003, AJ 125, 2029\\
Oasa Y., 2003, in proc IAU Symposium 211 'Brown Dwarfs', p91, ed. E.L. Martin, ASP,
  San Francisco\\
O'Dell C.R., Wen Z. 1994, ApJ 436, 194\\
O'Dell C.R., Yusef-Zadeh F, 2000, AJ, 120, 382\\
Petr M.G., Coude du Foresto V., Beckwith S.V.W., Richichi A., McCaughrean M.J. 
  1998, ApJ 500, 825\\
Pinfield D.J., Dobbie P.D., Jameson R.F., Steele I.A., Jones H.R.A., Katsiyannis A.C.
  2003, MNRAS 342, 1241\\
Potter D., Martin E.L., Cushing M.C., Baudoz P., Brandner W., Guyon O., 
  Neuhauser R. 2002, ApJ 567, L133\\
Prosser C.F., Stauffer J.R., Hartmann L., Soderblom D.R., Jones B.F., 
  Werner M.W., McCaughrean M.J. 1994, ApJ 421, 517\\
Rees M.J. 1976, MNRAS 176, 483
Reid I.N., Gizis J.E., Kirkpatrick J.D., Koerner D.W. 2001, AJ 121, 489\\
Reiners A., Basri G. 2005, in proc 'The 13th Cambridge Conference on Cool Stars,
  Stellar Systems and the Sun' (in press)\\ 
Reipurth B., Clarke C. 2001, AJ 122, 432\\
Riddick F.C., Roche P.F., Lucas P.W., 2005, in proc. 'IMF@50', Kluwer Academic Publishers,
  Astrophysics and Space Library series, eds. E. Corbelli, F. Palla, and H. Zinnecker,
  (in press)\\
Rieke G.H., Lebofsky M.J. 1985, ApJ, 288, 618\\
Robberto M., Soderblom D.R., O'Dell C.R., Stassun K.G., Hillenbrand L.A., Simon, M.,, 
  Feigelson E.D., Najita J., Stauffer J., Meyer M. + 12 authors 2004, AAS 205, 117.03\\
Scally A., Clarke, C. 2001, MNRAS 325, 449\\
Scally A., Clarke, C. 2002, MNRAS 334, 156\\
Silk J. 1977, ApJ 214, 152\\
Simon M. 1997, ApJ 482, L81\\
Slesnick C.L., Hillenbrand L.A., Carpenter J.M. 2004, ApJ 610, 1045\\
Stapelfeldt K.R., Krist J.E., Menard F., Bouvier J., Padgett D.L., Burrows C.J. 
  1998, ApJ 502, L65\\
Sterzik M.F., Durisen R.H., 1998 A\&A 339, 95\\
Tokunaga A.T., 2000, in Allen's Astrophysical Quantities; ed. A.Cox, AIP Press, New York\\
Volk K., Blum R., Walker G., Puxley P., 2003, IAU Circ., 8188, 2\\
Walker C., Wood K., Lada C.J., Robitaille T., Bjorkman J.E., Whitney B. 
  2004 ApJ 351, 607\\
Whitworth A.P., Zinnecker H. 2004, A\&A 427, 299\\

\pagebreak
\onecolumn
\begin{center}
Table 1 - Observational details.
\begin{tabular}{lcccc} \hline
Dataset & FWHM$^a$ & Integration & Time on source & Origin of\\ 
& (arcsec) & time (s) & (s) & zero point\\ \hline
Field 1: K & 0.51 & 20 & 3140 & Hillenbrand \& Carpenter (2000)\\
Field 1: H & 0.46 & 30 & 3240 & 2MASS 2nd Release \\
Field 1: J & 0.41 & 60 & 6900 & 2MASS 2nd Release \\
Field 2: K & 0.55 & 60 & 3600 & 2MASS 2nd Release \\
Field 2: H & 0.59 & 60 & 2820 & 2MASS All Sky \\
Field 2: J & 0.58 & 120 & 6960 & 2MASS All Sky \\
Field 3: K & 0.36 & 60 & 3480 & Gemini standard \\
Field 3: H & 0.59 & 60 & 3600 & 2MASS All Sky \\
Field 3: J & 0.53 & 120 & 8040 & 2MASS All Sky \\
\end{tabular}
\end{center}
\vspace{-2cm}
\small Notes: 
(a) This column gives the full width half maximum of point sources in the coadded datasets.
\normalsize

\vspace{1cm}
\begin{center}
Table 2 - Source catalogue.
\begin{tabular}{llllccccccc} \hline
Source  & Coordinate & RA$^a$ (J2000) & Dec (J2000) & K$^b$ & H & J & Flag$^c$ & Origin$^d$\\ 
Number & based ID &  & & &  & & & of Flux\\ \hline
   1 & 154-747  &  35  15.390 &  27  47.11 &  11.06 &  12.41 &  14.89 &     & 0,0,0 \\
   2 & 222-745  &  35  22.192 &  27  44.84 &  14.30 &  15.33 &  16.11 &     & 0,0,0 \\
   3 & 034-745  &  35   3.450 &  27  44.54 &  16.09 &  -1.00 &  -1.00 &     & 0,0,0 \\
   4 & 213-739  &  35  21.279 &  27  38.78 &  12.11 &  12.81 &  13.46 &     & 0,0,0 \\
   5 & 152-738  &  35  15.211 &  27  37.71 &  14.40 &  15.73 &  16.92 &     & 0,0,0 \\
   6 & 138-737  &  35  13.818 &  27  36.69 &  11.42 &  11.90 &  12.75 &     & 0,0,0 \\
   7 & 071-736  &  35   7.147 &  27  36.41 &  18.44 &  -1.00 &  -1.00 &     & 0,0,0 \\
   8 & 208-736  &  35  20.829 &  27  36.20 &  16.56 &  17.45 &  18.28 &  HJ & 0,0,0 \\
\end{tabular}
\end{center}
\vspace{-2cm}
\small Notes: 
(a) The RA and Dec are given in minutes and seconds east of 05h 00m 00s and arcminutes and 
arcseconds south of -5 degrees. Estimated precision is 0.5 arcseconds.\\
(b) Sources not detected in a passband have the following dummy values: 0.00 for undetected
sources; -1.00 for sources off the edge of the data mosaic; -2.00 for saturated sources
not seen in other surveys; -3.00 for binaries unresolved in some passbands.\\
(c) Highly uncertain fluxes ($>0.2$ mag) are indicated by letters in this column
denoting the passband(s) with uncertain flux.
(d) Sources in general have fluxes drawn from one or more datasets, indicated separately
for the K, H and J fluxes in that order. The source is indicated by the following key. 
0: this survey. 1: Lucas et al.(2001); 2: the 2MASS All Sky Survey; 3: the 2MASS 
2nd Incremental Data Release; 4: Hillenbrand \& Carpenter (2000); 5: Muench et al.(2002).  
\normalsize

\vspace{5cm}
\begin{center}
Table 3 - Binaries detected by Gemini and Subaru 
\begin{tabular}{lcccc}
ID$^a$ & RA$^b$ & Dec & Separation (``) & Comments\\ \hline
\multicolumn{4}{c}{Gemini sources} \\ \hline
268, 271 & 05 35 07.65 & -5 24 00.7 & 0.49 & \\
277, 278 & 05 35 09.68 & -5 23 55.9 & 0.48 & \\
288, 291 & 05 34 59.32 & -5 23 32.9 & 0.54 & \\
375, 376 & 05 35 06.20 & -5 22 12.3 & 0.39 & \\
221, 222 & 05 35 16.19 & -5 24 56.4 & 0.51 & \\
213, 216 & 05 35 05.01 & -5 25 00.9 & 0.72 & 216 is a brown dwarf candidate\\
265$^c$ &  05 35 08.23 & -5 24 03.2 & $\sim 0.40$ & Saturated source.\\ \hline
\multicolumn{4}{c}{Subaru sources} \\ \hline
217-147 & 05 35 21.7 & -5 21 47 & $\sim 0.6$ & saturated source.\\
205-330 & 05 35 20.5 & -5 23 30 & $\sim 0.6$ & saturated source.\\
184-427 & 05 35 18.4 & -5 24 27 & 0.39  &  \\
163-210 & 05 35 16.3 & -5 22 10 & 0.40  &  \\
129-135 & 05 35 12.9 & -5 21 35 & 0.67  &  \\
231-501 & 05 35 23.1 & -5 25 01 & 0.66  & near $\theta_2$ Ori\\
154-225 & 05 35 15.4 & -5 22 25 & 0.67  & \\
183-438 & 05 35 18.3 & -5 24 38 & 0.67  & \\ 
\end{tabular}
\end{center}
\small Notes:\\ 
(a) Source IDs from Table 2 are given for the Gemini binary members.
Coordinate based designations are used for the Subaru binaries.\\
(b) J2000.0 coordinates are given for the primary. \\
(c) Source 265 is a binary which is listed singly in Table 2. Separate 
photometry was not possible since both components were saturated and barely
resolved.\\
\normalsize
\twocolumn

\end{document}